\documentclass[aps,showpacs,superscriptaddress,twocolumn]{revtex4}

\usepackage{amsmath,bm}
\usepackage{graphicx}
\usepackage{epsfig}

\begin{document}

\newcommand{\vk}{{\vec k}}
\newcommand{\vK}{{\vec K}}
\newcommand{\vb}{{\vec b}}
\newcommand{{\vp}}{{\vec p}}
\newcommand{{\vq}}{{\vec q}}
\newcommand{\vQ}{{\vec Q}}
\newcommand{\vx}{{\vec x}}
\newcommand{\beq}{\begin{equation}}
\newcommand{\eeq}{\end{equation}}
\newcommand{\half}{{\textstyle \frac{1}{2}}}
\newcommand{\gton}{\stackrel{>}{\sim}}
\newcommand{\lton}{\mathrel{\lower.9ex \hbox{$\stackrel{\displaystyle<}{\sim}$}}}
\newcommand{\ee}{\end{equation}}
\newcommand{\ben}{\begin{enumerate}}
\newcommand{\een}{\end{enumerate}}
\newcommand{\bit}{\begin{itemize}}
\newcommand{\eit}{\end{itemize}}
\newcommand{\bc}{\begin{center}}
\newcommand{\ec}{\end{center}}
\newcommand{\bea}{\begin{eqnarray}}
\newcommand{\eea}{\end{eqnarray}}
\newcommand{\beqar}{\begin{eqnarray}}
\newcommand{\eeqar}[1]{\label{#1} \end{eqnarray}}
\newcommand{\pleft}{\stackrel{\leftarrow}{\partial}}
\newcommand{\pright}{\stackrel{\rightarrow}{\partial}}

\newcommand{\eq}[1]{Eq.~(\ref{#1})}
\newcommand{\fig}[1]{Fig.~\ref{#1}}
\newcommand{\eff}{ef\!f}
\newcommand{\alphas}{\alpha_s}

\renewcommand{\topfraction}{0.85}
\renewcommand{\textfraction}{0.1}
\renewcommand{\floatpagefraction}{0.75}

\title{A theory of jet shapes and cross sections:
from hadrons to nuclei}

\date{\today  \hspace{1ex}}

\author{Ivan Vitev}
\affiliation{Los Alamos National Laboratory, Theoretical Division,
MS B238, Los Alamos, NM 87545, USA}
\author{Simon Wicks}
\affiliation{Department of Physics, Columbia University, 538 West 120-th Street,
New York, NY 10027, USA}
\author{Ben-Wei Zhang}
\affiliation{Los Alamos National Laboratory, Theoretical Division,
MS B238, Los Alamos, NM 87545, USA}
\affiliation{Institute of Particle Physics, Hua-Zhong Normal University,
Wuhan, 430079, China}

\begin{abstract}

For jets, with great power comes great opportunity. The unprecedented
center of mass energies available at the LHC open new windows on the
QGP:
we demonstrate that jet shape and jet cross section measurements become
feasible as a new, differential and accurate test of the underlying  QCD
theory.  We present a first step in understanding these shapes and
cross sections
in heavy ion reactions. Our approach allows for detailed simulations of
the experimental acceptance/cuts that help isolate jets in such 
high-multiplicity environment. It is demonstrated for the first time that
the pattern of stimulated gluon emission can be correlated with a
variable quenching of the jet rates
and provide an approximately model-independent approach to determining
the characteristics of the medium-induced bremsstrahlung spectrum.
Surprisingly, in
realistic simulations of  parton propagation through the QGP we find a
minimal increase in the mean jet radius even for large jet attenuation.
Jet broadening is manifest in the tails of the energy distribution away
from the jet axis and its quantification requires high statistics
measurements that will be possible at the LHC.

\end{abstract}

\maketitle

\section{Introduction}

Fast partons propagating in a hot/dense nuclear medium are expected to lose
a large fraction of their energy~\cite{Wang:1991xy}. In fact, the stopping
power of strongly-interacting matter for color-charged particles has, by far,
the largest experimentally established effect: the attenuation of the cross
section for final-state observables of large mass/momentum/energy.
This jet quenching mechanism has been used to successfully  explain the
strong suppression of the hadron spectra at large transverse momentum
observed  in nucleus-nucleus collisions at the Relativistic Heavy
Ion Collider (RHIC)~\cite{Gyulassy:2003mc,:2008cx}. There is
mounting evidence that the quark-gluon plasma (QGP)-induced
quenching can be disentangled from other nuclear effects even at
the much lower  Super Proton Synchrotron (SPS) center of mass
energy~\cite{Aggarwal:2007gw}. To calculate parton energy loss in the QGP several
theoretical approaches have been developed~\cite{BDMPS-Z-ASW,GLV,HT,AMY}.
While tackling the same basic problem, they use different assumptions
for the boundary conditions (initial/final quark or gluon virtuality), make
different  approximations for the parent parton and radiative gluon
kinematics, and  treat  differently the interaction between the jet+gluon
system with the medium (differentially vs on average). For discussion
see~\cite{Wicks:2008ta}.

At present, most measurements of hard processes are limited to single
particles and particle correlations, which are only the leading fragments
of a jet. There is general agreement on the physics that controls inclusive
particle suppression in the QGP and the experimental methodology
of determining $R_{AA}(p_T)$ (or $I_{AA}(p_{T_1},p_{T_2})$)~\cite{Gyulassy:2003mc}.
Thus, for any  particular combination of radiative/collisional energy loss
evaluation and its phenomenological application to leading particle quenching
the  QGP density can be determined  with $20-25\%$  ``statistical''
accuracy~\cite{Adare:2008cg}. An inherent limitation of this approach
is that while fits to the data can, not surprisingly, always be
performed~\cite{Bass:2008rv} they do not resolve the staggering order of
magnitude "systematic" discrepancy in the extracted medium properties.
Furthermore, focusing on quantities that can be constrained with little
ambiguity from the measured rapidity entropy density in heavy ion collisions
distracts from issues such as the  approximations that go into theoretical
energy loss derivations and their application to systems where
more often  than not these initial assumptions are
violated~\footnote{ Some examples include the assumption
that the coupling constant $g_s\ll 1$, while $g_s \approx 2$, or
the assumption  that $L/\lambda_g \rightarrow \infty \; (\geq 10)$,
while $L/\lambda_g \approx 5$. Many jets undergo even fewer
interactions in the QGP since they originate in the periphery
of the interaction region.}. In searching for experimental measurements
which  can  pinpoint the framework for energy loss calculations
that is applicable to heavy ion reactions, complex
multi-particle correlations may not be optimal. They are very
sensitive to non-perturbative effects/fragmentation and the  modeling
effort cannot be systematically improved due to the violation of factorization
for highly exclusive observables~\cite{Collins:2007nk}. It is, therefore,
critical  to  find alternatives that  accurately reflect the energy flow in
strongly-interacting  systems and have a more direct connection
to the underlying quantum chromo-dynamics (QCD) theory.

The intra-jet energy distribution and the related cross section for
jets in the case of heavy ion reaction closely match the criteria
outlined above. The high rate of hard probes at the LHC and the
large-acceptance calorimetry, see e.g.~\cite{D'Enterria:2007xr},
will enable these accurate measurements. It should be noted that
proof-of-principle measurements of jet cross sections have become
possible at RHIC~\cite{Salur:2008hs}, but significantly better
statistics will be required to quantify the QGP effects on jets. In
this paper we study the magnitude of these modifications in Pb+Pb
collisions at $\sqrt{s}=5.5$~TeV at LHC. We demonstrate that a
natural generalization of leading particle suppression to jets,
\begin{equation}
R_{AA}^{\text{jet}}(E_T; R^{\max},\omega^{\min}) =
\frac{ \frac{d\sigma^{AA}(E_T;R^{\max},\omega^{\min})}{dy d^2 E_T} }
{ \langle  N_{\rm bin}  \rangle
\frac{d\sigma^{pp}(E_T;R^{\max},\omega^{\min})}{dy d^2 E_T}  } \; ,
\label{RAAjet}
\end{equation}
is sensitive to the nature of the medium-induced energy loss. The
steepness of the final-state differential spectra amplifies the observable effect
and the jet radius $R^{\max}$ and the minimum particle/tower energy
$p_{T\, \min} \approx \omega^{\min}$ provide, through the
evolution of $R_{AA}^{\text{jet}}(E_T; R^{\max},\omega^{\min})$ at any fixed
centrality, experimental access to the QGP response to  quark and gluon
propagation.

In the following discussion of jet shapes and jet cross sections in p+p and
A+A collisions we stay as close as possible to an analytic theoretical
approach. Thus, we are able to unambiguously connect the
non-perturbative QCD effects, medium properties and the induced
bremsstrahlung spectrum to experimental jet observables. Determination
of baseline jet shapes and their generalization to finite momentum
acceptance cuts builds upon the work of Seymour~\cite{Seymour:1997kj}.
We refer the reader to~\cite{Seymour:1997kj,Ellis:2007ib}
for discussion of the complications in defining a jet and the related topic
of jet-finding algorithms. To make the discussion simpler, we will assume
that the complications of the different definitions can be subsumed into
an $R_{sep}$ parameter, as described in Appendix \ref{app:params}. 
Once a jet axis and all of the jet
particles / calorimeter towers ``i'' have been identified, the ``integral jet
shape''  is defined as:
\begin{equation}
 \Psi_{\text{int}}(r;R) = \frac{\sum_i (E_T)_i \Theta (r-(R_{\text{jet}})_i)}
{\sum_i  (E_T)_i  \Theta(R-(R_{\text{jet}})_i)}
\end{equation}
where $r,R$ are Lorentz-invariant opening angles, $R_{ij} =
\sqrt{(\eta_i-\eta_j)^2 + (\phi_i-\phi_j)^2}$, and $i$ represents
a sum over all the particles in this jet. $\Psi_{\text{int}}(r;R)$ is the
fraction of the total energy of a jet of radius $R$ within a sub-cone
of radius $r$.  It is automatically normalized so that
$\Psi_{\text{int}}(R;R) = 1$.
To move from the integrated to the differential jet shape, we define:
\begin{equation}
 \psi(r;R) = \frac{d\Psi_{\text{int}}(r;R)}{dr}  \;.
\end{equation}
This is the angular density of jet energy (remembering that the
appropriate 3D representation would be $\psi^{vis}(r;R) =
\frac{1}{2\pi r}\psi(r;R)$). Understanding the many-body QCD theory
behind jet shape calculations will naturally lead to understanding
the attenuation of jets in reactions with heavy nuclei.

This article is organized as follows: in Section~\ref{sec:shapespp}, we outline
a calculation of the jet shape in nucleon-nucleon (N+N) collisions using
the framework of perturbative QCD. We compare this calculation to existing
Tevatron data and investigate the jet shapes at LHC energies. A brief discussion
of final-state QGP-induced radiative energy loss in the GLV formalism is given
in Section~\ref{sec:shapesMed}. We prove that the cancellation of small-angle near-jet
axis bremsstrahlung persists to all orders in the correlation between the multiple
scattering centers and provide details of its numerical evaluation. The fully
differential distribution of the energy lost by a hard parton is also shown.
In Section~\ref{sec:shapesTot} we present results for the medium-modified
jet shapes and cross sections as a function of the jet cone radius, $R$, and
the experimental $p_T$ cut, $\omega_{\min}$, and discuss a simple energy sum
rule. We demonstrate the connection between the characteristic properties of
the QGP-induced gluon radiation and the variable suppression, at the same impact
parameter $b$, of jet rates, the modulation of the mean jet radius and the
enhancement in the ``tails''  of the intra-jet energy flow distribution.
A summary and conclusions are presented in Section~\ref{sec:summary}.
Appendix~\ref{app:cross} shows a calculation of the baseline jet cross
sections at the LHC and an estimate of the accuracy with which these
cross sections and jet shapes can be measured with nominal first-year integrated
luminosities in p+p and A+A reactions. In Appendix~\ref{app:params}  we study
the influence of the different perturbative and non-perturbative contributions
to the jet shape in hadronic collisions. Finally, Appendix~\ref{app:dz}
contains a discussion of a double differential measure of energy flow in jets
and its connection to particle angular correlation measurements, currently
conducted at RHIC.

\section{Jet shapes in 'elementary' p-p collisions}
\label{sec:shapespp}

In the process of advancing perturbative QCD theory to  many-nucleon
systems, the final-state experimental  observables  should be first
understood
in the simpler p+p reactions.

\subsection{Theoretical considerations}
\label{subsec:theory}

\subsubsection{Leading order results}

In the introduction, we defined the central quantity of our study,
the differential jet shape for parton ``a'',  $\psi_a(r;R)$. As
in~\cite{Seymour:1997kj},  the starting point of the calculation
is the leading  order parton splitting: a suitable separation
of physical time scales enables the separation of the
calculation into production and jet showering.
The QCD splitting functions $P_{a \rightarrow bc}(z)$  give the
distribution of the large fractional lightcone momenta
(or approximately the energy fractions) of the fragments relative
to the parent parton, $z$ and $1-z$ respectively. To lowest order,
recalling that $\psi_a(r;R)$ describes the energy  flow $\propto z$,
we can write:
\begin{equation}
   \psi_a(r;R) = \frac{ d  \Psi_{\text{int},a}(r;R) }{ dr } =
\sum_b \frac{\alpha_s}{ 2 \pi }
\frac{2}{r} \int_{z_{min}}^{1-Z} dz  \,  z P_{a \rightarrow bc}(z) .
\label{kernel}
\end{equation}
In Eq.~(\ref{kernel})  $ r = (1-z) \rho $ is related to the opening
angle $\rho$ between the final-state partons.


In ``elementary'' p+p collisions the inclusion of soft particles
($ z_{min} \approx 0 $) in theoretical calculations is not a bad
approximation. Even in this case, however,
there are intrinsic limitations on the minimum particle/calorimeter
tower
$p_T$ or $E_T$, related,  for example, to  detector acceptance.
In heavy ion reactions, especially  for the most interesting case of
central collisions,
there is an enormous background of soft particles related to the bulk
QGP properties. Jet studies will likely require  minimum particle energy
$ > 1-2$~GeV at RHIC and even more stringent cuts at the LHC.
Furthermore, control over $z_{min}$ can provide detailed information
about the properties of QGP-induced bremsstrahlung. Further kinematic
constraints on the values of $z$  arise since both the resulting partons
must be within an angular distance $R$ of the original jet axis,
$r<R,\, rz/(1-z)<R$. In this case they are identified with the jet. If
not, they are identified as two  separate jets. For a cone-based
algorithm, the relative separation $R_{sep}R$  (as opposed to just the
distance  from the original jet axis) is an additional criterion:
$\rho<R_{sep}R$. We find:
\begin{eqnarray}
  Z &=& \max \left\{ z_{min},   \frac{r}{r+R} \right\} \;  \mbox{ if }
\;
r < (R_{sep} - 1)R  \;,   \\
   Z  &=& \max \left\{ z_{min},  \frac{r}{R_{sep}R} \right\} \;
\mbox{ if } \;
r > (R_{sep} - 1)R  \; .
\label{Zpar}
\end{eqnarray}
Carrying out the integration in Eq.~(\ref{kernel}) we arrive at the LO
jet shape functions for quarks and gluons:
\begin{eqnarray}
  \psi_q(r) &=& \frac{C_F \alpha_s}{2\pi} \frac{2}{r}  \left(
  2 \log \frac{1-z_{min}}{Z}  \right. \nonumber \\
&& \left. - \frac{3}{2} \left[ (1-Z)^2 -z_{min}^2 \right]  \right)  \; ,
\label{psiLO1} \\
  \psi_g(r) &=& \frac{C_A \alphas}{2\pi} \frac{2}{r}
\left( 2 \log \frac{1-z_{min}}{Z}     \right. \nonumber \\
&&  \left. -  \left(  \frac{11}{6}
- \frac{Z}{3} + \frac{Z^2}{2} \right) (1-Z)^2  \right. \nonumber \\
&& \left. + \left( 2 z_{min}^2 -\frac{2}{3} z_{min}^3
+ \frac{1}{2} z_{min}^4  \right) \right) \nonumber \\
     &&  + \frac{T_R N_f \alphas}{2\pi} \frac{2}{r}   \left(
  \left( \frac{2}{3} - \frac{2Z}{3} + Z^2 \right) (1-Z)^2 \right.
  \nonumber \\
&& \left.   -\left(  z_{min}^2 - \frac{4}{3} z_{min}^3 + z_{min}^4
\right)  \right) \;.
\label{psiLO2}
\end{eqnarray}
In the $z_{min} \rightarrow 0$ limit Eqs.~(\ref{psiLO1})
and~(\ref{psiLO2})
reduce and we recover the previously known results~\cite{Seymour:1997kj}.
There is an implicit `plus-prescription' in these results when
calculating
moments of physical quantities, as we have
not considered the virtual corrections in the forward direction.
Hence, the result is not applicable for $r=0$ and does not have the
correct normalization when integrated. However, the shape
is reflective of final-state parton splitting and, when needed,  for
the leading order calculation one may apply a cutoff for small $r$ and
normalize via first-bin subtraction.

In contrast to the case of $ e^+ + e^- $ annihilation, hadronic
scattering is accompanied by copious initial-state radiation (ISR)
that can fall within the jet cone. While the contribution of the
ISR is small for small values of $ r/R $,  it gives an
essential contribution at larger angles. A simple estimate
based on a dipole radiation and the kinematics of the hard parton - soft
gluon coincidence within a cone~\cite{Seymour:1997kj}, similar for both quark
and gluon jets, yields:
\begin{equation}
\psi_i(r) = \frac{C \alphas}{2 \pi} 2 r \left( \frac{1}{Z^2}
   - \frac{1}{(1-z_{min})^2} \right) \;,
\label{loIS}
\end{equation}
Again, the `plus-prescription' to account for the $r=0$ point
is not explicitly shown. In Eq.~(\ref{loIS})  $C \simeq C_F \approx
C_A/2$.

The leading order calculation is most appropriate for the rare,
hard splittings of a very high momentum jet. The use of a running
coupling improves the numerical results,  providing larger  weight for
softer events. We employ a running $\alpha_s$ evaluated at the largest
$k_T$ in the problem, $\mu = r (1-Z) E_T$ for the jet splitting and
$\mu = (1-Z)E_T$ for the initial state radiation~\cite{Seymour:1997kj}.

\subsubsection{Resummation - all orders and multiple emission}

As $r\rightarrow 0$, in the collinear limit of parton splitting, the
leading
order contributions to the jet shape diverge, see
Eqs.~(\ref{psiLO1}),~(\ref{psiLO2})
and~(\ref{loIS}).  In fact, all orders in the perturbative expansion
diverge,
including powers of $\log r$ in the form $\alphas^n \log^{2n-1}r$.
With plentiful parton showering, it becomes increasingly less
likely
that any particular quark and gluon will be coincident with the jet
axis. Quantitatively, this is described by a Sudakov form factor.
The energy density at small angles is dominated by the hard parton
in the splitting. If there is a splitting that leaves the hard parton
at an angle $r_1$, a subsequent splitting at $r_2 < r_1$ will not
contribute to the energy density at $r_2$. Multiple independent
splitting
follows a Poisson distribution, hence, the probability of energy flow
at an angle less  than $r$ is exponentially suppressed by the
integrated probability at angles greater than $r$, i.e.:
\begin{eqnarray}
  P(<r) &=& \exp (-P_1(>r)) \\
        &=& \exp \left( -\int_r^R dr' \, \psi_{\rm coll}(r')  \right)
\;.
\label{poisson}
\end{eqnarray}
This only applies for soft emissions which do not take away (much)
momentum,
i.e. at leading log accuracy. Improvement can be obtained at modified
leading
log accuracy (MLLA), when the running of the coupling constant is
included
in $P_1$. We don't take  other (e.g. recoil or kinematic constraints)
effects in evaluating the Sudakov form factor in the soft collinear
approximation.

The resummed $\psi_{\rm resum}(r) = \frac{d}{dr}P(<r)$ and we carry
out the integration in
Eq.~(\ref{poisson}), including the running $\alpha_S(rE_T)$, to obtain
modified leading
logarithmic accuracy (MLLA). Note that $ \alpha_s(\mu) = 1/(2\beta_0
\log\frac{\mu}{\Lambda_{QCD}}) $, with $4 \pi \beta_0 = b_0 = \frac{11}
{3} C_A -
\frac{4}{3} T_R N_f$. First, take the small $r$ limit in
Eqs.~(\ref{psiLO1}),
(\ref{psiLO2}) and~(\ref{loIS}), keeping terms $ \propto 1/r $ and
$ \propto 1/r \log(1/r) $.
Based on  $Z = \max (z_{min},r/R)$, in this limit we have two
kinematic domains.
For $r>z_{min}R$ the results are similar to the known case of no
acceptance cut-off
and reduce to the known results if $z_{min} = 0$:
\begin{eqnarray}
&&P_q(r>z_{min}R)= \exp\left( 2C_F \log\frac{R}{r} f_1\left(2 \beta_0
\alpha_s
\log\frac{R}{r} \right)  \right.  \nonumber \\
&& \hspace*{1.cm}-  \left[ \frac{3}{2}C_F  - CR^2  - c_q^>(z_{min})
\right]
\label{GrQ}   \nonumber \\
&& \hspace*{1.cm}
\left. \times \, f_2\left(2 \beta_0 \alpha_s \log\frac{R}{r} \right)
\, \right)  \;, \\
&&P_g(r>z_{min}R)= \exp\left( 2C_A \log\frac{R}{r} f_1\left(2 \beta_0
\alpha_s
\log\frac{R}{r} \right)  \right.  \nonumber \\
&& \hspace*{1.cm}
-  \left[ \frac{1}{2} b_0  - CR^2   - c_g^>(z_{min})  \right]
\label{GrG}   \nonumber \\
&& \hspace*{1.cm}
\left. \times \, f_2\left(2 \beta_0 \alpha_s \log\frac{R}{r} \right)
\, \right) \; .
\end{eqnarray}
It is useful to employ the same notation as in~\cite{Seymour:1997kj}
and facilitate
the comparison to the case of no kinematic cuts: $f_1(x)=\log(1-x)/
(2\pi\beta_0)$, and
$ f_2(x) =  ( 1 - \log(1-x)/x ) / (2\pi\beta_0) $. The $z_{min}$-
dependent corrections
are isolated as follows:
\begin{eqnarray}
c_q^>(r>z_{min}R;z_{min}) & = &  2 C_F \log(1-z_{min})  \nonumber \\
&& \hspace*{-1.5cm}
+ \frac{3}{2}C_F z_{min}^2 \;, \qquad \\
c_g^>(r>z_{min}R;z_{min}) & = &  2 C_A \log(1-z_{min})  \nonumber \\
&& \hspace*{-1.5cm}
+  C_A  \left( 2 z_{min}^2 -\frac{2}{3} z_{min}^3  + \frac{1}{2}
z_{min}^4  \right)
\nonumber \\ && \hspace*{-1.5cm}
- T_RN_f \left(  z_{min}^2 - \frac{4}{3} z_{min}^3 + z_{min}^4
\right) \;.  \qquad
\end{eqnarray}
When $r< z_{min} R$ the integration in Eq.~(\ref{poisson}) has to be
split in two regions:
$r^\prime \in (r,z_{min}R) $ and $r^\prime \in (z_{min}R, R) $. The
second integral is
trivially obtained from the case that was just considered above, Eqs.
(\ref{GrQ})
and (\ref{GrG}),   with the substitution
$r=z_{min} R$.  When combined with the first integral, it yields:
  \begin{eqnarray}
&&P_q(r<z_{min}R)= P_q(r>z_{min}R; r=z_{min}R) \nonumber \\
&& \hspace*{1.cm}   \times \exp\left( -\left[ \frac{3}{2}C_F -
c_q^<(z_{min})    \right]
\right. \nonumber \\
&& \hspace*{1.cm} \left.  \times f_2\left(2 \beta_0 \tilde{\alpha}_s
\log\frac{z_{min}R}{r} \right)  \, \right)  \;, \qquad \\
&&P_g(r<z_{min}R)=P_g(r>z_{min}R;r=z_{min}R) \nonumber \\
&&  \hspace*{1.cm}   \times \exp\left( -\left[ \frac{1}{2}b_0  -
c_g^>(z_{min}) \right]
\right. \nonumber \\
&& \hspace*{1.cm} \left.  \times
   f_2\left(2 \beta_0 \tilde{\alpha}_s \log\frac{z_{min}R}{r} \right)
\, \right) \; . \qquad
\end{eqnarray}
Here, we denote by $\tilde{\alpha}_s = \alphas(z_{min}RE_T)$, as
opposed to
$ \alpha_s = \alphas(RE_T)$, and:
\begin{eqnarray}
c_q^<(r<z_{min}R;z_{min}) & = &  2 C_F \log\left(\frac{1-z_{min}}
{z_{min}}\right)
\nonumber \\
&& \hspace*{-1.5cm} + 3 C_F z_{min} \;, \qquad \\
c_g^<(r<z_{min}R;z_{min}) & = &  2 C_A  \log\left(\frac{1-z_{min}}
{z_{min}}\right)
  \nonumber \\
&& \hspace*{-1.5cm}
+  C_A  \left( 4z_{min} - z_{min}^2 + \frac{2}{3} z_{min}^3 \right)
\nonumber \\ && \hspace*{-1.5cm}
- T_RN_f \left( 2 z_{min} - 2z_{min}^2 + \frac{4}{3}z_{min}^3   \right) \;. \qquad
\end{eqnarray}
Note that the Sudakov form factors evaluated here will regulate any
collinear
divergence present in $\psi_{\rm coll}$.

\subsubsection{Power corrections - an estimate of non-perturbative
effects}
Inclusion of the running coupling constant under the momentum transfer
integrals
yields contributions from regions in which $Q \sim \Lambda_{QCD}$
or lower, i.e. there is a fundamental non-perturbative contribution to
all of the integrals. An estimate of these power correction effects with
finite acceptance gives the following result:
\begin{eqnarray}
\psi_{PC}(r) &=& \frac{2C_R}{2\pi} \frac{2}{r} \frac{Q_0}{rE_T}
\Bigg( \bar{\alpha_0}'(Q_0,k_{min}) - \alphas(\mu)  \nonumber \\
&&   \left. - 2 \beta_0 \alphas(\mu)^2
\left( 1 + \log \frac{\mu}{Q_0}  \right)  \right) \nonumber \\
&& + \frac{2C_R}{2\pi} \frac{2}{r} \frac{k_{min}}{rE_T} \Bigg(
\alphas(\mu) \nonumber \\
&& \left. + 2 \beta_0 \alphas(\mu)^2 \left( 1 +
\log \frac{\mu}{k_{min}}  \right)  \right) \; ,
\label{powerFS}
\end{eqnarray}
where $C_R = C_F, C_A$ for quarks or gluons, respectively.
In Eq.~(\ref{powerFS}) $k_{min}=z_{min}rE_T$ and
$\mu$ is the renormalization scale.
The term  $ \propto \bar{\alpha_0}'(Q_0,k_{min})$ depends on
the parametrized non-perturbative contribution, defined as:
\begin{eqnarray}
\bar{\alpha_0}'(Q_0,k_{min}) = \frac{1}{Q_0}\int_{k_{min}}^{Q_0} dk \,
\alphas(k) \; ,
\label{eqn:alphas_0}
\end{eqnarray}
with $Q_0$ representing the non-perturbative scale.
In our numerical calculation we use
\begin{eqnarray}
\bar{\alpha_0}'(2\, \text{GeV},0) = 0.52 \,\, , \,\,\,
\bar{\alpha_0}'(3\, \text{GeV},0) = 0.42 \,\,
\label{eqn:Q0}
\end{eqnarray}
from Ref.~\cite{Dokshitzer:1995zt,Seymour:1997kj}
and a  parametrization of the strong coupling constant at small
momentum transfer given in Ref.~\cite{Fischer:2002hna}.
The terms  $\propto \alpha_s(\mu), \, \alpha^2_s(\mu), $ come from
subtracting  the perturbative component in the non-perturbative
region~\cite{Dokshitzer:1995zt}.  Finally, the term  $\propto k_{min}$
results from the introduction of finite acceptance, $z_{min}$.

A similar expression is derived for initial-state radiation:
\begin{eqnarray}
  \psi_{i,PC}(r) &=& \frac{2C}{2\pi} 2r \frac{Q_0}{E_T}
\Bigg( \bar{\alpha_0}'(Q_0,k_{min}') - \alphas(\mu)  \nonumber \\
&&   \left. - 2 \beta_0 \alphas(\mu)^2
\left( 1 + \log \frac{\mu}{Q_0}  \right)  \right) \nonumber \\
&& + \frac{2C_R}{2\pi} 2r \frac{k_{min}'}{E_T} \Bigg(
\alphas(\mu) \nonumber \\
&& \left. + 2 \beta_0 \alphas(\mu)^2 \left( 1 +
\log \frac{\mu}{k_{min}'}  \right)  \right) \; ,
\end{eqnarray}
where $k_{min}'=z_{min} E_T$, and $C \simeq C_F \approx C_A/2$.
At lower jet energies the power corrections are, in fact, sizable
even at large $r$. This suggests that if there is deviation between the
theoretical results and the  experimental data it
may be largely due to incomplete consideration of non-perturbative
effects. We stress that the generalization of power corrections to
finite acceptance implies that these should be taken into account only
for
$z_{min} r E_T < Q_0$ or $z_{min} E_T < Q_0$ for final-state and
initial-state
radiation, respectively.

\subsubsection{Total contribution to the jet shape}

As indicated  before, the resummed jet shape at small $r/R$ is
evaluated as
$\psi_{\text{resum}}(r)=\frac{d}{dr} P(r) = \psi_{\text{coll}}(r)P(r)
$. Taking
all contributions to the jet shape  and ensuring that there is no double
counting at small $r/R$ to $ {\cal{O}}(\alpha_s)$ we find:
\begin{eqnarray}
\psi(r)&=&\psi_{\text{coll}}(r)\left( P(r)-1\right) +
\psi_{\text{LO}}(r) + \psi_{i,\text{LO}}(r)
\nonumber \\ && + \psi_{\text{PC}}(r)
+ \psi_{i,\text{PC}}(r) \; ,
\label{totpsi}
\end{eqnarray}
On the right-hand-side of Eq.~(\ref{totpsi}) the first term comes
from  Sudakov resummation  with  subtraction of the leading  $1/r$,
$ (1/r) \log (1/r) $ contribution at small $r/R$ to avoid double
counting with the fixed order component of the differential jet shape.
The second and  third terms represent the leading-order contributions in the
final-state  and the initial-state.  The last two terms represent  the effect
of  power corrections. In a full calculation the relative quark and gluon
fractions $f_q+f_g = 1$  are also needed: $ \psi (r;E_T) = f_q(E_T,\sqrt{s})
\psi_q(r;E_T) + f_g(E_T,\sqrt{s})   \psi_g(r;E_T)$. These fractions
are calculated in  Appendix~\ref{app:cross} alongside the demonstration
of the feasibility of jet cross section and differential jet shape
measurements at $\sqrt{s}=5.5$~TeV.

If the resummed part completely dominates the area under $\psi(r)$
in Eq.~(\ref{totpsi}), i.e.
the power corrections and the fixed order result affect only the
large $r/R$ ``tails'', the theoretically calculated differential jet shape is
properly normalized. In reality, this is not the case and the first
correction, ${\cal O}(\alpha^2_s)$, arises  from  $\psi_{\rm coll}(r) (P(<r) -1)
= \psi_{\rm coll}(r) ( 1 + C_R \alpha_s (\cdots) + \cdots - 1 )$.
It will be larger for gluon jets when compared to quark jets, $C_A$ vs $C_F$,
and for lower transverse energies. The normalization can them be ensured
via
\begin{equation}
\psi(r) \rightarrow \psi(r) + {\rm Norm} \times \psi_{\rm coll}(r) \ln (P(<r)) \;,
\label{normalize}
\end{equation}
where ${\rm Norm}$ is determined numerically.  We stress that  to achieve a robust
theoretical description of  the differential jet shape all contributions to
$\psi(r)$ from  Eq.~(\ref{totpsi})  should be included. In Appendix ~\ref{app:params}
we elucidate their relative strength using numerical examples.
We also investigate the dependence of the shape on
$R_{sep}$ and the non-perturbative scale $Q_0$.

\subsection{Comparison to the Tevatron data}

\begin{figure}[!t]
\centerline{\hbox{
\epsfig{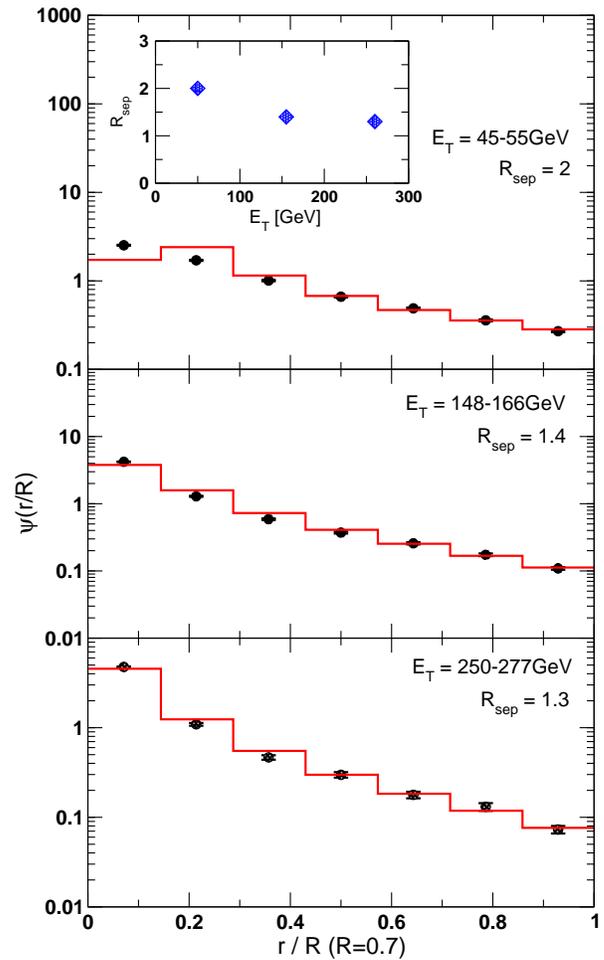}
}}
\caption{(Color online) Comparison of numerical results
from our theoretical calculation to experimental data on
differential jet shapes at $\sqrt{s}=1960$~GeV by
CDF II~\cite{Acosta:2005ix}. Insert shows the $E_T$
dependence of $R_{sep}$.
}
\label{fig:CDF}
\end{figure}

In Fig.~\ref{fig:CDF} we show comparison of the theoretical
model for the jet shape,  Eq.~(\ref{totpsi}), to
the experimental measurements in $p+\bar{p}$ collisions at
$\sqrt{s}=1960$~GeV at Fermilab from Run II (CDF II)~\cite{Acosta:2005ix}.
Our numerical results include all contributions from leading order,
resummation and power corrections with $Q_0 = 2$~GeV. The insert
shows the  variation of the parameter $R_{sep}$ with the
transverse energy of the jet. At high jet $E_T$ our
theoretical model gives very good descriptions of the large $r/R$
experimental data with $R_{sep} = 1.3-1.4$. For  $E_T=45-55$~GeV
the largest meaningful value  $R_{sep} = 2$ can describe the data
fairly well, except at very small $r/R$ region. Extended discussion
of the various contributions to the differential jet shape is
given in Appendix ~\ref{app:params}.

We note that for $r/R \ll 1$ and a  large gluon jet fraction
in conjunction with moderate $E_T \leq 50$~GeV  there is still
deviation between the data and the theory, e.g. the top panel of
Fig.~\ref{fig:CDF}. This is likely related to the need for significant
corrections, Eq.~(\ref{normalize}), to ensure the proper normalization
of $\psi(r/R)$.  Such corrections, in turn, point to NLO effects, a
possible breakdown of our soft collinear jet splitting approximation
for the Sudakov resummation and non-perturbative effects. Note that
even Monte Carlo event generators have to be tuned to describe this
data~\cite{Acosta:2005ix}. For the purpose of our manuscript the
deficiencies in this specific part of phase space are not essential
since, as we will see in Section~\ref{sec:shapesTot}, the
experimental signatures of jet propagation in the QGP are
most pronounced in the complementary  $r/R \sim 1$ domain.
One simply has to keep in mind that the description of  $r/R<0.25$
$\psi(r/R)$ in the  vacuum allows for further theoretical improvement.

\subsection{Predictions for the LHC}

\begin{figure}[!t]
\centerline{\hbox{
\epsfig{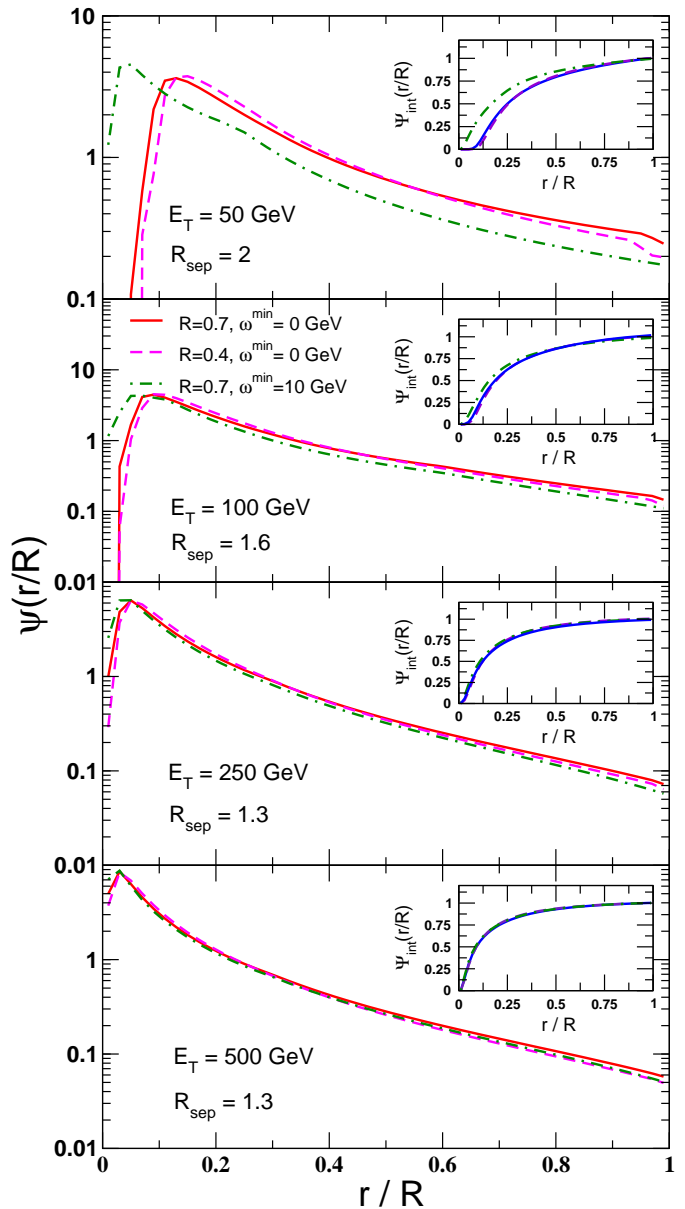}
}}
\caption{(Color online) Numerical results for the differential jet shapes
in p+p collisions at $\sqrt{s}=5.5$~TeV at the LHC. Solid lines
represent jet shapes  with
$R=0.7,\, \omega^{\min}=0$~GeV, dashed lines stand for jet shapes with
$R=0.4,\; \omega^{\min}=0$~GeV, and dashed-dotted
lines are for jet shapes with $R=0.7,\; \omega^{\min}=10$~GeV. The inserts show
integrated jet shapes $\Psi_{\rm int}(r;R)$.
}
\label{fig:LHCshapes}
\end{figure}

We employ the theoretical model that describes the CDF II data
and apply the same transverse energy-dependent $R_{sep}$ parameter
to obtain  predictions for the LHC at $\sqrt{s}=5.5$~TeV. The emphasis
here is to produce a baseline in $p+p$ reactions for comparison to the
full in-medium jet shape in  $Pb+Pb$ collisions. The essential
difference in going from the Tevatron to the LHC is in the production
of hard jets. At the higher collision energy  we observe a
greater  contribution from gluon jets relative to quark jets,  e.g.
Fig.~\ref{sigmajet1} in Appendix~\ref{app:cross}. Therefore, for the
same $E_T$, jets at the LHC are expected to be slightly wider than
at the Tevatron.

\begin{figure}[!ht]
\epsfig{file=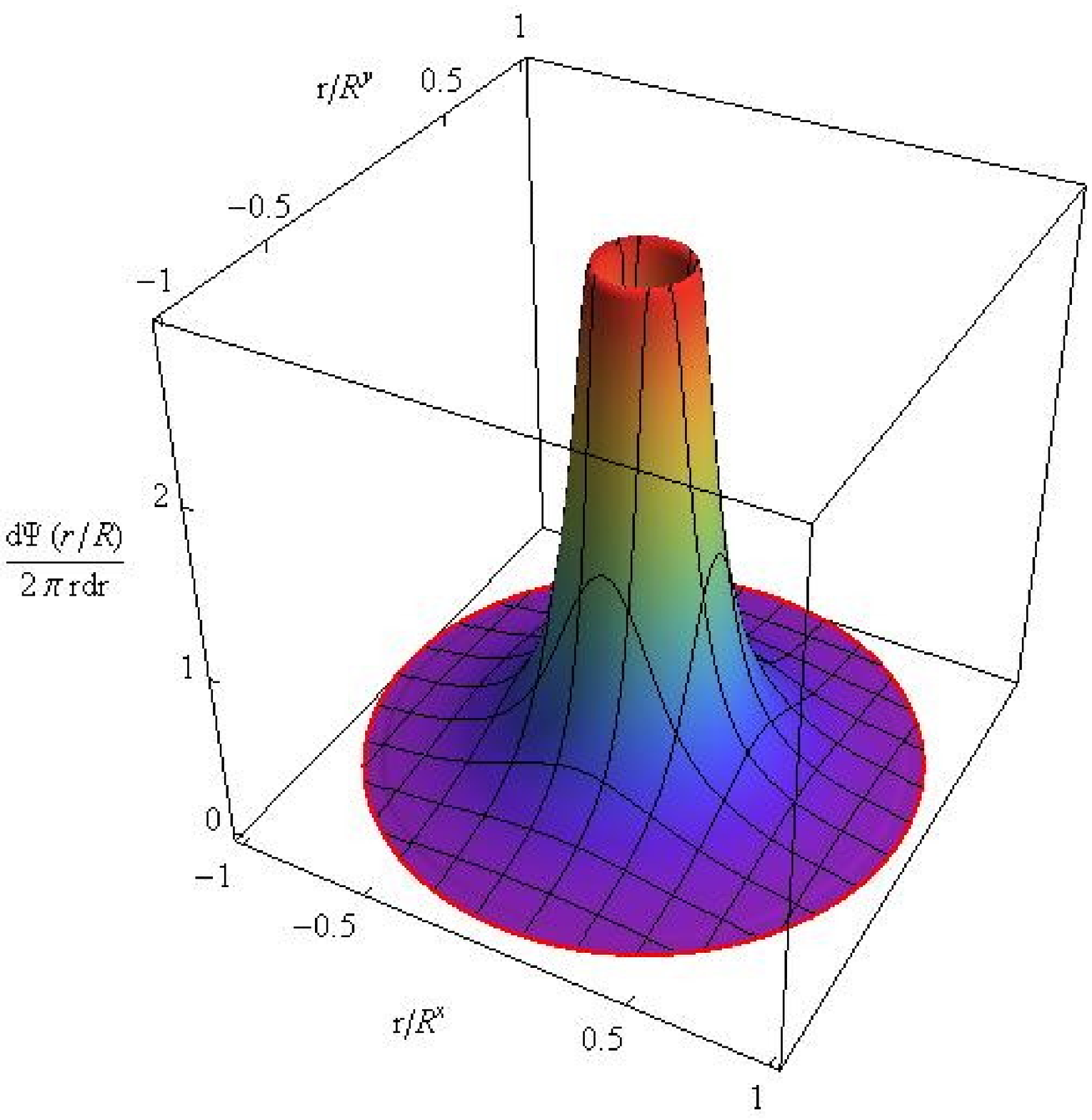,height=2.5in,clip=,angle=0}\\
\epsfig{file=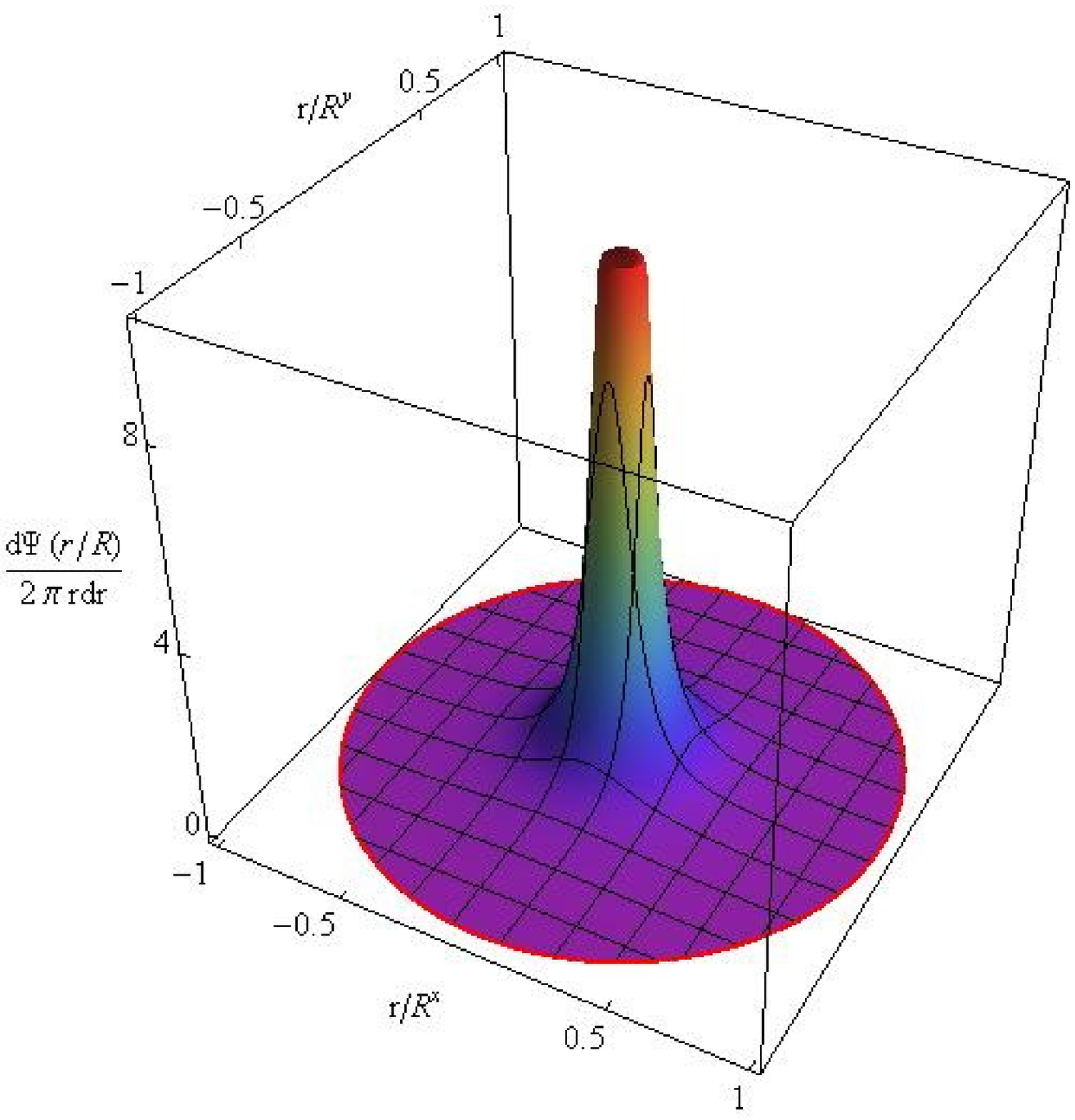,height=2.5in,clip=,angle=0}\\
\epsfig{file=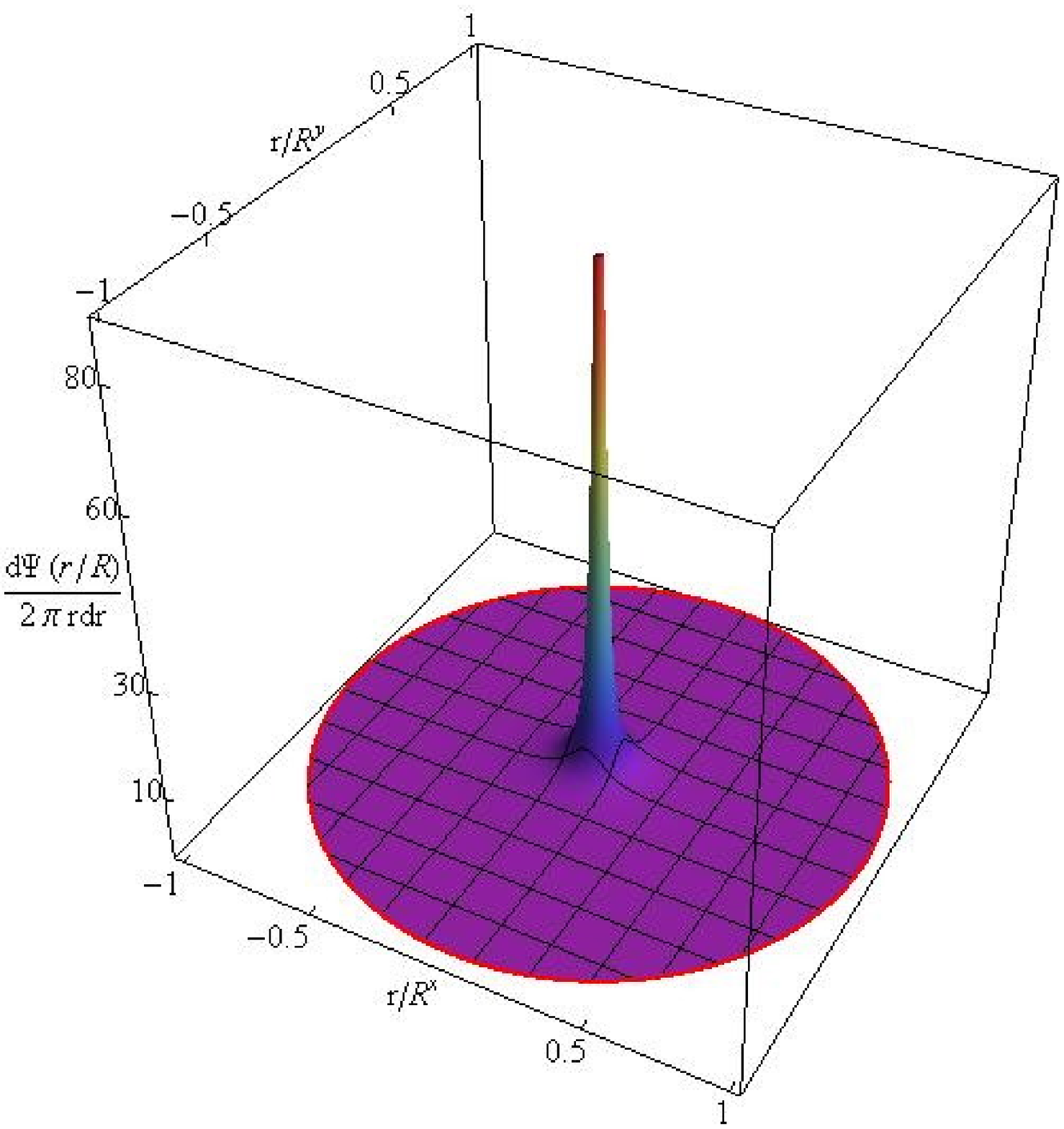,height=2.5in,clip=,angle=0}\\
\caption{(Color online) 3D plot of the differential jet shapes
at three different jet energies $E_T=20$~GeV (top panel),
$E_T=100$~GeV (middle panel), and $E_T=500$~GeV (bottom panel)
with  $R=0.7,\, \omega^{\min}=0$~GeV
in p+p collisions with $\sqrt{s}=5500$~GeV at the LHC.
From  low jet energy to high jet energy, jet shape becomes much steeper.
}
\label{fig:LHCshapes-3D}
\end{figure}

Figure~\ref{fig:LHCshapes}  shows our numerical results for the jet shape for
four different energies $E_T = 50, \; 100,\; 250\;, 500$~GeV and two cone
radii $R=0.7,\; 0.4$
in $p+p$ collisions at  $\sqrt{s}=5.5$~TeV at LHC. An interesting
observation is that, when plotted against the relative opening angle
$r/R$,  these shapes are self-similar, i.e. approximately independent
of the absolute cone radius $R$. One of the main theoretical developments
in this paper is the analytic approach to studying finite detector
acceptance effects or experimentally imposed low momentum cuts. In
Fig.~\ref{fig:LHCshapes} this is illustrated via the selection of
$z_{min} = p_{T\, \min} / E_T = 0.2, \; 0.1,\;  0.04$ and 0.02   ($p_{T\, \min}
= \omega^{\min} = 10$~GeV). Eliminating the soft partons naturally
leads to a  narrower  branching  pattern. However, for this effect to be
readily observable $10-20\%$ of the jet energy, going into soft particles,
must be missed. Thus, even with $p_{T\, \min} \sim$few GeV cuts in Pb+Pb
collisions at the LHC aimed at reducing or eliminating the background
of bulk QGP particles that accidentally fall within the jet cone, the
alteration of $\psi(r/R)$ is expected to be small.
We  also studied the integral jet shape $\Psi_{\rm int}(r/R)$, shown in the
inserts of Fig.~\ref{fig:LHCshapes}, as a tool for identifying kinematic
and dynamic effects on jets~\cite{Salgado:2003rv}.  Only when large
differences exist between two $\psi(r/R)$ for $r/R < 0.4$ these will be
reflected in the integral jet shape. If the differences are pronounced
in the $r/R > 0.4$ region, as is typically the case for heavy ion
reactions,  $\Psi_{\rm int}(r/R)$ will be practically insensitive to the
QGP effects on jet propagation.

It is important to also note in Fig.~\ref{fig:LHCshapes} that there is
a dramatic change in the differential jet shape in going from small to
large transverse energies even in $p+p$ reactions. This is best demonstrated on
a 3D-plot, where the volume of the jet cone is normalized to unity. Examples of
$\frac{d\psi(r/R)}{2\pi r dr}$ for
$E_T = 20,\; 100,\; 500$~GeV are given in Fig.~\ref{fig:LHCshapes-3D}.
We have used $R=0.7$ and $ \omega^{\min}=0 $. It is obvious that
development of detailed  theoretical models and their validation
against experimental data in  nucleon-nucleon collisions are necessary
before any credible conclusions about the modification of the QCD jets in
the QGP medium can be drawn.

 \section{Medium-Induced Contribution to the Jet Shape}
\label{sec:shapesMed}

The principal medium-induced contribution to a jet shape comes from
the radiation pattern of the fast quark or gluon, stimulated by their
propagation and interaction in the QGP. There is a simple heuristic
argument which allows one to understand how interference and coherence
effects in QCD amplify the difference between the energy distribution
in a vacuum jet and  the in-medium jet shape~\cite{Vitev:2008jh}.
Any destructive effect on the integral average parton energy
loss  $\Delta E^{rad}$, such as the Landau-Pomeranchuk-Migdal effect,
can be traced at a differential level to the attenuation or full
cancellation  of the collinear,  $k_T \ll \omega$,  gluon
bremsstrahlung:
\begin{eqnarray}
&& \Delta E^{rad}_{\rm LPM \; suppressed} \Rightarrow
\frac{dI^g}{d\omega}(\omega \sim E)_{\rm LPM \; suppressed} \nonumber \\
&& \Rightarrow  \frac{dI^g}{d\omega d^2 k_T}
(k_T \ll \omega )_{\rm LPM \; suppressed} \;, \;
\label{largeang}
\end{eqnarray}
and we indicate the parts of phase space where the modification of
the incoherent $ {dI^g} / {d\omega d^2 k_T} $ is most effective.

Indeed, detailed derivation of the coherent inelastic parton
scattering regimes
in QCD was given in~\cite{GLV}. In all cases, the origin of the
LPM suppression can be traced  to the cancellation of the collinear
bremsstrahlung. The destructive quantum interference is most prominent
for
final-state radiation, where the large-angle gluon bremsstrahlung was
originally
discussed in Ref.~\cite{Vitev:2005yg} to first order in opacity. Even
though to
carry out  realistic simulations to higher orders in opacity with full
geometry will require computational power beyond what is currently
available,
we first present an analytic proof that a cone-like pattern of
medium-induced emission persists to all orders in the correlations
between
multiple scattering centers (elementary emitters) and we focus on the
case of immediate interest: light quark and gluon jets and final-state
bremsstrahlung. Generalization to massive partons can easily be
achieved,
see e.g.~\cite{Djordjevic:2003zk}, but it is important  to note that the
effect of a heavy quark mass versus the jet energy depends on the
coherent
scattering regime~\cite{GLV}.

\subsection{ Radiative energy loss in the GLV formalism }

In our calculation we will use the GLV formalism of expanding the
medium-induced radiation in the correlations between multiple scattering
centers~\cite{GLV}.  We first recall the definitions of the Hard,
Gunion-Bertsch and Cascade propagators in terms of the gluon transverse
momentum  ${\bf k}$  and the transverse momentum transfers from the
medium
${\bf q}_i$:
\begin{eqnarray} {\bf H}&=&{{\bf k} \over {\bf k}^2 }\; ,
  {\bf C}_{(i_1i_2 \cdots i_m)}={({\bf k} - {\bf q}_{i_1} - {\bf
q}_{i_2}
- \cdots -{\bf q}_{i_m} ) \over ({\bf k} - {\bf q}_{i_1} -
{\bf q}_{i_2}- \cdots -{\bf q}_{i_m} )^2 } \;,
\nonumber \\[1.ex]
{\bf B}_i &= &{\bf H} - {\bf C}_i \; , {\bf B}_{(i_1  \cdots i_m
)(j_1 \cdots i_n)} = {\bf C}_{(i_1  \cdots j_m)} - {\bf C}_{(j_1
\cdots j_n)}\;\; ,\nonumber \\
\label{Eq:GB}
\end{eqnarray}
The relevant inverse gluon formation times can be written as:
\begin{eqnarray}
[ \tau^f_{(i_1i_2 \cdots i_m)} ]^{-1}=  \omega_{(i_1i_2 \cdots i_m)}
&=& [ k^+ |{\bf C}_{(i_1i_2 \cdots i_m)}|^2 ]^{-1} \; . \;  \qquad
\end{eqnarray}
For final-state radiation, the intensity spectrum reads:
\begin{widetext}
\beqar
k^+ \frac{dN^g(FS)}{dk^+ d^2 {\bf k} } &=& \frac{C_R \alpha_s}{\pi^2}
\sum_{n=1}^{\infty}  \left[ \prod_{i = 1}^n  \int
\frac{d \Delta z_i}{\lambda_g(z_i)}  \right]
\left[ \prod_{j=1}^n \int d^2 {\bf q}_j \left( \frac{1}{\sigma_{el}
(z_j)}
\frac{d \sigma_{el}(z_j) }{d^2 {\bf q}_j}
-  \delta^2 ({\bf q}_j) \right)    \right] \nonumber \\
&& \times \;  \left[ -2\,{\bf C}_{(1, \cdots ,n)} \cdot
\sum_{m=1}^n {\bf B}_{(m+1, \cdots ,n)(m, \cdots, n)}
\left( \cos \left (
\, \sum_{k=2}^m \omega_{(k,\cdots,n)} \Delta z_k \right)
-   \cos \left (\, \sum_{k=1}^m \omega_{(k,\cdots,n)}
\Delta z_k \right) \right)\; \right]  \;,   \qquad
\eeqar{full-final}
\end{widetext}
where $\sum_2^1 \equiv 0$ and  ${\bf B}_{(n+1, n)} \equiv {\bf B}_n$
are understood. In the case of final-state interactions,
$z_0 \approx 0$ is the point of the initial hard scatter and $z_L = L$
is
the extent of the medium.  The path ordering  of the
interaction points, $z_L > z_{j+1} > z_j > z_0$, leads to the
constraint
$\sum_{i=1}^n \Delta z_i  \leq  z_L $. One implementation of this
condition would be $\Delta z_i \in [\, 0,z_L -\sum_{j=1}^{i-1}
\Delta z_j \, ]$
and it is implicit in Eq.~(\ref{full-final}).

There is an obvious limit of the GLV radiative spectrum when
$L \gg \lambda_g \gg \tau_f $, where $\lambda_g$
is the mean free path of the gluon in a hot QGP.  Here,  the
contributions of
the $\cos(\cdots)$ terms vanish after integration over the
unobserved  ${\bf q}_{i}$ or
$\Delta z_i$ due to rapid oscillation. It is easy to see  in this limit
for n=1 that,
\begin{eqnarray}
k^+\frac{dN^g}{dk^+} &=& \frac{C_R \alpha_s}{\pi^2}
\left\langle  \frac{L}{\lambda_g}  \right\rangle
  \int d^2{\bf k} \int d^2{\bf q}_{1}\, \nonumber \\
& & \times  \left\langle \frac{1}{\sigma_{el}}
\frac{d \sigma_{el} }{d^2 {\bf q}_1} \right\rangle
\,  [{\bf C}_1^2 - {\bf H}^2 + {\bf B}_1^2]\; .
\label{Eq:opacity-1}
\end{eqnarray}
In the very high energy limit $E \rightarrow \infty$, leading
to  large ${\bf k}$ phase space,  a change of variables ${\bf k}
\rightarrow
{\bf k-q_1}$  shows that  the first two terms in
Eq.~(\ref{Eq:opacity-1}),
cancel, leading to an incoherent Bertsch-Gunion gluon emission in a
hot QGP medium with $\langle n \rangle =  \frac{L}{\lambda_g}$. By
direct
inspection one can see that the $n\geq 2$ terms do not contribute. In
fact, it is
easy to verify that for any bremsstrahlung  regime, initial-state,
final-state
and no hard scattering, this limit holds~\cite{GLV}.  More generally,
in this limit it can be shown that the reaction operator
$\hat{R}  \rightarrow 0$. Naturally, for finite jet energies there
will be
corrections when $k^+ {dN^g}/{dk^+ d^2 {\bf k} }$ is evaluated
numerically with actual kinematic bounds~\cite{Wicks:2008ta}.
\vspace*{.1cm}

\subsection{Collinear radiation in GLV formalism }

While the example given above illustrates that limits can be
imposed and taken in the GLV results, such limits are artificial in
that the formation time of the gluon at the emission vertex spans
$\tau_f \in (0,\infty)$. The Reaction Operator approach~\cite{GLV},  
i.e. the GLV formalism, is not an
approach of averages: it compares differentially $\tau_f $ to the
separation between the scattering centers. For example, even when
${\bf k} \rightarrow 0$ the formation time can be small or large,
depending on the momentum transfers for the medium. Let us investigate
this case in more detail: we note that  $k^+ \approx 2 \omega$  and
${\bf k} \approx r \omega \hat{n}$, where $r$ is the  angle relative
to the jet axis. Here, $\hat{n}$ is a unit vector transverse to the
jet axis which defines the azimuthal angle $\phi$ of gluon emission.
Using the results of Eq.~(\ref{full-final}),
the 2D $(\phi, r)$ angular distribution of gluons at n-th order in
the correlated scattering expansion reads:

\begin{widetext}

\beqar
\lim_{ r \rightarrow 0 } \frac{ \omega dN^g}
{d\omega d \phi  d r }
&\propto &  \omega  \left[ \prod_{j=1}^n \int d^2 {\bf q}_j
  \left( \frac{1}{\sigma_{el}(z_j)}
\frac{d \sigma_{el}(z_j) }{d^2 {\bf q}_j}
- \delta^2 ({\bf q}_j) \right)    \right]  \;
\frac{ {\bf q}_1 + \cdots + {\bf q}_n}
{({\bf q}_1 + \cdots + {\bf q}_n)^2}
  \cdot  \sum_{m=1}^n  \omega r \bigg(
\frac{  {\bf q}_{m+1} + \cdots + {\bf q}_n }
{ ({\bf q}_{m+1} + \cdots + {\bf q}_n)^2 }
  \nonumber \\ &&
-  \;  \frac{ {\bf q}_m + \cdots + {\bf q}_n}
{({\bf q}_m + \cdots + {\bf q}_n)^2}  \bigg)
   \times \; \left(  \cos \left( \sum_{k=2}^m
\frac{ ( {\bf q}_k + \cdots + {\bf q}_n)^2 }{2\omega}
\Delta z_k    \right)   -  \cos \left( \sum_{k=1}^m
\frac{ ( {\bf q}_k + \cdots + {\bf q}_n)^2 }{2\omega}
\Delta z_k   \right) \; \right)  \; . \; \qquad
\eeqar{finalfull-det}

\end{widetext}
Here, we have already set $r=0$ where possible. We can use  this
general notation as long as we clarify certain special cases: for $m =
n$ we
have $\cos [ ({\bf q}_{n+1} + {\bf q}_n)^2  \Delta z_{n+1} /2 \omega ]
\equiv 1$.  For the transverse propagators and $m$ we have
$ \omega r ({\bf q}_{n+1} + {\bf q}_n)/
({\bf q}_{n+1} +  {\bf q}_n)^2 \equiv \hat{n} $.
It is know that the leading $ n = 1 $  contribution to
final-state medium-induced radiation  leads to  $ \lim_{ r \rightarrow
0 }
{ \omega dN^g} / {d\omega d \phi  d r } = 0$~\cite{Vitev:2005yg}.
Our goal is to show  that this result is general and holds to any
order in the expansion. Its implications are  that there is very little
overlap between the techniques used to compute the ``vacuum'' and
medium-induced contributions to the jet shape. A general proof requires
demonstration of the absence of unprotected divergences for any set
of momentum transfers $\{ {\bf q}_i \}$, finiteness of the
momentum transfer integrals as ${\bf q}_i \rightarrow \infty$ and
a mechanism that kills the small-angle contribution.

\begin{enumerate}

\item
We first look at the large ${\bf q}_i$ limit. The transverse
propagator contribution itself in Eq.~(\ref{finalfull-det}) behaves
as $\sim 1/{\bf q}_i^2$. Furthermore, irrespective of the small
  ${\bf q}_i$ behavior of the momentum transfer distribution from
the medium, for large momentum transfers the  collisional cross
section is suppressed by the Rutherford  $\sim 1/{\bf q}_i^4 $
behavior, ensuring the finiteness of the integrals.

\item
Next, we examine the potential singularity as
$| {\bf q}_1 + \cdots + {\bf q}_n |  \rightarrow 0$.
The difference in the LPM interference  terms in this limit goes as
$ {\cal O}( ({\bf q}_1 + \cdots + {\bf q}_n)^2  )$ and for the
most problematic transverse propagator term (m=1) even as
$ {\cal O}( ({\bf q}_1 + \cdots + {\bf q}_n)^4  )$.
In summary, not only is there no divergence in this case,
but the integrand in Eq.~(\ref{finalfull-det}) vanishes.

\item
We now collect the interference phases associated with problematic
propagators as
$| {\bf q}_k + \cdots + {\bf q}_n |  \rightarrow 0$,
$1 < k \leq n$. Expanding for a small net transverse momentum
sums we find that the singularity is canceled:
\begin{eqnarray}
&&
     \bigg[ \sin \sum_{j=2}^{k-1}
\frac{( {\bf q}_j + \cdots + {\bf q}_n )^2}{2\omega} \Delta z_j
\\
&&  - \sin \sum_{j=1}^{k-1}
\frac{( {\bf q}_j + \cdots + {\bf q}_n )^2}{2\omega} \Delta z_j \bigg]
\frac{( {\bf q}_k + \cdots + {\bf q}_n )^2}{2\omega} \Delta z_k  \; .
\nonumber
\end{eqnarray}
Actually, the lack of singularities persists also away from the
small $r$ limit.

\item

With the integrand well behaved and all integrals finite we see that
the phase space factor $r$ in the numerator is sufficient to ensure
vanishing medium-induced bremsstrahlung contribution at the center of
the
jet. It is assisted by partial cancellation from angular integrals of
the type  $\int {\bf q}_i \cdot  {\bf q}_j \, f({\bf q}_i, {\bf q}_j) \,
d \phi_{ij} $. It is only for the special case of
$\hat{n} \cdot ( {\bf q}_1 + \cdots + {\bf q}_n ) $ where
the antisymmetric integrand under ${\bf q}_i \rightarrow  - {\bf q}_i $
for all $ i$ fully ensures the vanishing zero-angle radiative
contribution.

\end{enumerate}
This completes our proof that at any order in opacity
\beqar
\lim_{ r \rightarrow 0 } \frac{ \omega dN^g_{\rm med}}
{d\omega d \phi  d r }  = 0 \;.
\eeqar{zero}
Numerical  simulations, using  Monte-Carlo techniques,
confirm independently  that  ${dI^g}/{d\omega d^2 {\bf k}}$
vanishes as  ${\bf k} \rightarrow 0$  ~\cite{Wicks:2008ta}.

\subsection{Numerical methods and  QGP properties}
\label{subsec:QGP}

Results relevant to the LHC phenomenology  are calculated using full
numerical evaluation  of the medium-induced contribution to the observed
jet shapes and the modification of the in-medium jet cross sections. Jet
production,
being rare in that $ \sigma(E_T > E_{T\; \min}) T_{AA}(b)  \ll 1$,
follows
binary collision  scaling $ \sim d^2N_{\rm bin.}/d^2{\bf x}_\perp $.
In contrast,
the medium is distributed according to the number of participants
density
$ \sim d^2N_{\rm part.}/d^2{\bf x}_\perp $. Soft particles that carry
practically all of the energy deposited in the fire ball of
a heavy ion collision cannot  deviate
a lot from such scaling.  We take into account longitudinal
Bjorken expansion since transverse expansion leads to noticeable
corrections only in the extreme $\beta_T \rightarrow 1$
limit~\cite{Gyulassy:2001kr}. In our approach all relevant
finite time   and finite kinematics integrals, such as the
ones over the separation   between
the scattering centers $\Delta z_i = z_i - z_{i-1}$, the bremsstrahlung
gluon phase space $\Lambda_{QCD} < \omega < E_{\rm jet.}$,
$\Lambda_{QCD} < k_\perp < 2 \omega $~\footnote{This condition allows
for the deflection of the jet and can be also derived from the finite
rapidity
range constraint $ 0 < y_g < y_{\rm jet} $ for the emitted gluon},
and the
transverse
momentum transfers $0 < q_i < \sqrt{s/4} = \sqrt{m_D E_{\rm jet}/2 }$,
are done numerically~\cite{GLV}. In our simulation we generated
in-plane jets, $ \phi_{\rm jet} - \phi_{\rm reaction\; plane} = 0 $.
This
is of little importance in central Pb+Pb collisions (b=3 fm), where
the medium
effects on jet propagation are most pronounced, but in
semi-central (b=8 fm) and peripheral (b=13 fm) reactions this will
lead to smaller than average energy loss.

The evolving intrinsic momentum and length scales in the QGP expected to
be created at the LHC are determined as follows:  we first estimate the
QGP formation time $\tau_0 = 1 / \langle p_T \rangle = 0.23$~fm,
where $ \langle p_T \rangle  \approx 850$~MeV is obtained
from extrapolations to LHC energies made by
using Monte Carlo event generator results, fit to the CDF
collaboration data
from $ \sqrt{s} = 1.8 $~GeV $p+\bar{p} $ collisions~
\cite{Sjostrand:2006za}.
Here we account for the observed $\sim 25\%$ increase in the mean
transverse
momenta in going from $N+N$  to $A+A$ collisions at RHIC.  Gluons
dominate
the soft parton multiplicities at the LHC and their time- and
position-dependent
density can be related to charged hadron rapidity density in the Bjorken
expansion model~\cite{Markert:2008jc}:
\begin{equation}
\rho =  \frac{1}{\tau}  \frac{d^2 ( dN^g/dy)}{d^2 {\bf x}_\perp}
\approx  \frac{1}{ \tau } \frac{3}{2}
\left| \frac{d\eta}{dy} \right| \frac{d^2 (dN^{ch}/d\eta) }{d^2 {\bf
x}_\perp}
\;  .
\label{ydep}
\end{equation}
Here,  $dN^{ch}/d\eta = \kappa N_{\rm part.}/ 2$ with $\kappa \approx
9$ for
$\sqrt{s} = 5.5$~TeV.

Table~\ref{table-props} summarizes characteristics  of Pb+Pb
collisions at the LHC and initial QGP properties. An inelastic cross
section  $\sigma_{in} = 65$~mb has been used in an optical Glauber model
where necessary.
\begin{table}[!b]
\begin{center}
\begin{tabular}{||c||c|c|c||}
\hline
    Class   &  Central
&    Mid-central  & Peripheral   \\
\hline \hline
    $b\; [fm]$  &  3 & 8  &  13  \\
\hline
   $N_{part}$  &  361  &   165 &  18  \\
\hline
   $dN^g/dy$  &  2800  &  1278 &  137  \\
\hline
   $\langle  T(\tau_0)\rangle $ [MeV]   &  751  &   693  &  426  \\
\hline
   $\langle  m_D(\tau_0)\rangle $ [GeV]   &  1.89  &  1.73  &  1.07  \\
\hline
   $\langle  \lambda_g(\tau_0) \rangle $ [fm]   &  0.25  &   0.27 &
0.46  \\
\hline
\end{tabular}
\end{center}
\caption{  Summary of the relevant energy loss
parameters ind initial QGP properties
for central, semi-central and peripheral collisions at $
\sqrt{s}=5.5$~TeV
collisions at the LHC. }
\label{table-props}
\end{table}
Assuming local thermal equilibrium one finds:
\begin{equation}
T(\tau, {\bf x}_{\perp}) = \  ^3\!\sqrt{ \pi^2 \rho(\tau,{\bf
x}_{\perp}) / 16 \zeta(3) }\;, \tau > \tau_0 \; .
\label{tempdet}
\end{equation}
The Debye screening scale is given by $m_D = gT$, recalling that we work
in the
approximation of a gluon-dominated plasma and $N_f = 0$. The relevant
gluon
mean free path is easily evaluated:  $\lambda_g = 1/ \sigma^{gg} \rho$
with
$\sigma^{gg} = (9/2) \pi \alpha_s^2 / m_D^2$.
Note that in Table~\ref{table-props} the quoted initial mean
temperatures,
Debye screening
scales and gluon mean free paths are obtained as averages with the
binary collisions
weight $T_{AA}({\bf x}_\perp;b)$. In our evaluation we use  $g_s  =
2.5,\;
(\alpha_s = 0.5)$  to describe the scattering of the jet with the
medium  but the QGP-induced bremsstrahlung is calculated with  a
running $
\alpha_s(k_T)$ for emission vertex, similar to the MLLA approach for
the vacuum jet shapes.

Evaluation of the medium-induced energy loss and its contribution to
jet shapes is numerically expensive. Exact results have been obtained
only for $ E_{\rm jet} = 20, \; 100 $ and 500~GeV. We interpolate for
other
values of interest.

\subsection{Energy loss distribution}

A consistent energy loss theory provides complete information for the
differential
distribution of the lost $\Delta E_{rad}$, i.e. the bremsstrahlung
spectrum  in
Eq.~(\ref{full-final}). The main point that we  make here is that
this distribution is completely determined by the properties of the
QGP and
the mechanisms of energetic quark and gluon stopping in hot and dense
matter.
Therefore, selecting  different jet radii $R$ and $p_{T\; \min}$ of
the particles will
significantly alter both the jet shape and the amount of energy lost
by the hard
parton which can be recovered in the experimental measurement. In
contrast,
we  have seen in Fig.~\ref{fig:LHCshapes} that the jet shapes scale
approximately as
a function of  $r/R$, i.e. they are independent of the selection of
cone opening angle
$R$. The jet cross section weakly depends on $R$, unless $R
\rightarrow 0$. Finally,
$z_{min} = 0.1 - 0.2$ is necessary to noticeably alter the jet shape,
implying that
$\sim 20\% $ of the parent parton energy has to be missed via  $p_{T\;
\min}$ cuts
to observe significant effects on $ \psi_{vac}(r) $.

\begin{figure}[!t]
\centerline{\hbox{
\psfig{file=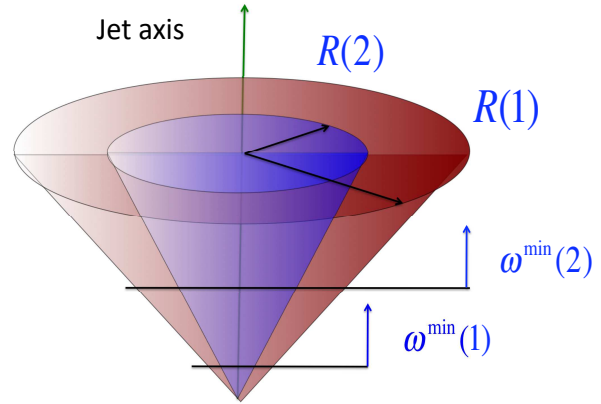,height=3.5in,width=2.8in,clip=,angle=-90}
}}
\caption{(Color online) Schematic illustration of the cone radius  $R$
and the  particle/tower $p_T$ / $E_T$ selection. The measured energy is
the one that comes from particles with  $p_T > \omega_{min}$ and
within $R$.
}
\label{fig:selectCone}
\end{figure}

\begin{figure}[!t] \hbox{
\epsfig{file=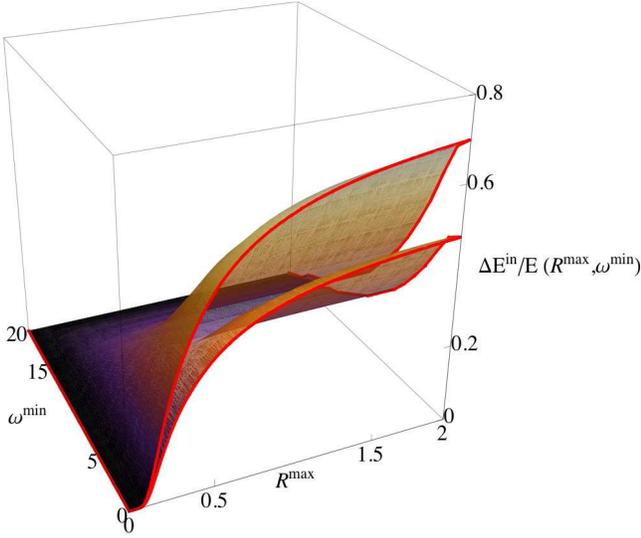,height=2.8in,width=3.50in,angle=0}
}
\caption{(Color online) 3D plot for the ratio of the energy that a
partons loses
inside a jet cone of opening angle $R^{\max}$  with  $\omega >
\omega^{\min}$ to the
total  parton energy. We have chosen a jet of energy $E_{jet} = 20$~GeV
in $b=3$~fm Pb+Pb collisions at LHC and varied the jet radius $R^{\max}
$ and
the acceptance cut $\omega^{\min}$. The upper surface is for a gluon jet
and the lower surface is for a quark jet.}
\label{fig:ERatio-3d}
\end{figure}

Experimentally, a clear strategy will  be to use the leverage arms
provided  by $R$ ($=R^{max}$  in the evaluation of the $\Delta E_{\rm
rad}$)
and $p_{T\, \min}$
($=\omega^{min}$  in the evaluation of the $\Delta E_{\rm rad}$) to
determine the
distribution of the lost energy. This is illustrated schematically in
Fig.~\ref{fig:selectCone}. Theoretically, the first quantity to be
calculated
is:
\begin{eqnarray}
  \frac{\Delta E^{in}}{E}(R^{\max},\omega^{\min})
= \frac{1}{E} \int_{\omega^{\min}}^E d\omega
  \int_0^{R^{\max} } dr  \frac{dI^g}{d\omega dr} (\omega,r) \; .
\nonumber \\
\label{def:out}
\end{eqnarray}
We present in Fig.~\ref{fig:ERatio-3d}  this fractional energy loss
for a quark jet and a gluon jet of energy $E_{jet} = 20$~GeV inside a
jet cone of radius $R^{\max}$ and with acceptance cut  $\omega^{\min}$.
Increasing $R^{\max}$ or decreasing $\omega^{\min}$ we will recover
more of the parent parton energy, lost via gluon bremsstrahlung.
We note that in Fig.~\ref{fig:ERatio-3d}  the mean energy loss was
calculated
as an average over the probability distribution
$P(\epsilon; E)$~\cite{Vitev:2005he}, reflective of multi-gluon
fluctuations:
\begin{equation}
\langle \epsilon \rangle = \left\langle \frac {\Delta E}{E} \right
\rangle = \int_0^1
d\epsilon \; \epsilon P(\epsilon; E) \;.
\end{equation}
For large fractional energy losses, such as in the illustrative example
of a  20 GeV jet in central $Pb+Pb$ collisions at the LHC,
$\Delta E_g/\Delta E_q$ is much smaller
than the asymptotic ratio $C_A/C_F = 9/4$ due to the kinematic
constraint
$\Delta E < E$~\cite{Vitev:2005he}.

\begin{figure}[!t]
\centerline{\hbox{
\epsfig{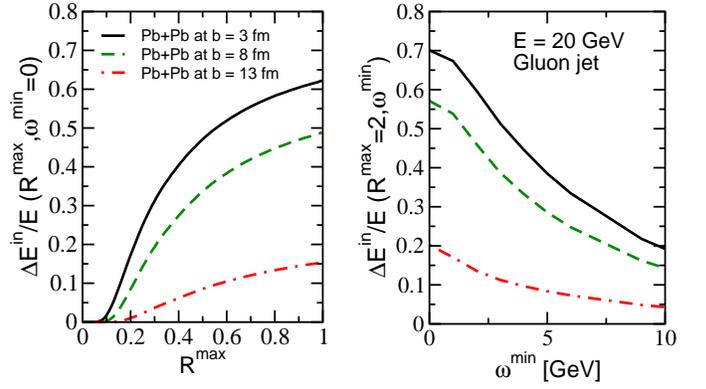}
}}
\caption{(Color online)  2D projections of Eq.~(\ref{def:out})
The left panel shows the fractional energy loss dependence on the jet
radius $R^{\max}$ ($\omega^{\min} = 0)$
and the right panel shows this dependence versus the  acceptance cut $
\omega^{\min}$
($R^{\max}= R^{\infty}$). A gluon jet of $E_{jet} = 20$~GeV
in $b = 3, \; 8, \; 13$~fm   $Pb+Pb$ collisions at LHC was used as an
example. }
\label{fig:ERatio-2d}
\end{figure}

The separate dependence of ${\Delta E^{in}(R^{\max},\omega^{\min})}/{E} $ on
the cone radius and the momentum acceptance cut is  more clearly
illustrated  in
Fig.~\ref{fig:ERatio-2d}. We show central, mid-central and peripheral
collisions,
impact parameters $b=3, \; 8, \; 13$~fm, respectively, in $Pb+Pb$
reactions
at LHC at nominal $\sqrt{s}$. We notice that, not surprisingly, the
ratio
$\Delta E^{in}(R^{\max},\omega^{\min})/E$ goes down at larger impact
parameters
because the energy loss of the jet decreases in peripheral collisions.
More importantly, at each impact parameter there is a variation of the
amount
of the bremsstrahlung energy, recovered in the cone. This is precisely
the variation that will map on the $R_{AA}^{\text{jet}}(R^{\max},
\omega^{\min})$ observable.
For example, in the limit of a very small opening angle and/or large
momentum
cut to eliminate the  QGP-induced radiation the suppression should
approximate
that of leading hadrons (up to differences arising from the possibly
softer particle
spectra due to fragmentation):
\begin{eqnarray}
R_{AA}^{jet}(R^{max}\rightarrow 0 \; {\rm and/or} \; \omega^{min}
\rightarrow E)
&=&  R_{AA}^{\rm leading \; parton}  \nonumber  \\
&\approx& R_{AA}^{h^\pm} \;.
\label{Rjet}
\end{eqnarray}
One can see that the typical choices, $R=0.4$ and $
\omega^{\min}=2$~GeV are a
good starting point to explore the variable quenching of jets.

\section{Tomography of Jets in Heavy Ion Collisions}
\label{sec:shapesTot}

The purpose of this Section is to relate the theory of jet propagation
in the QGP to experimentally measurable quantities.

\subsection{Experimental observables}

An essential ingredient that controls the relative contribution of
$\psi_{\rm vac.}(r/R)$ and $\psi_{\rm med.}(r/R) = (1/\Delta E_{\rm rad}) dI^g/dr$
to the observed differential jet shape in heavy ion reactions and  also
determines the attenuation of the jet cross sections is:
\begin{eqnarray}
f &\equiv& f\left(\frac{R}{R=R^\infty}, \frac{\omega^{\min}}{\omega^{\min}=0}\right)
\nonumber  \\
&=& \frac{\Delta E_{\rm rad}\left\{ (0,R);(\omega^{\min},E) \right\} }
{\Delta E_{\rm rad} \left\{ (0,R^\infty);(0,E) \right\} } \; ,
\end{eqnarray}
the {\em fraction} of the lost energy that falls within the jet cone, $r < R$,
and carried by gluons of $\omega > \omega^{\min}$ relative to the
total parton energy loss without the above kinematic constraints. If this
fraction is known  together with the probability distribution $P(\epsilon)$
for the parton energy loss  the medium-modified jet cross section per
binary $N+N$ scattering can be calculated as follows:
\begin{eqnarray}
\frac{\sigma^{AA}(R,\omega^{\min})}{d^2E_Tdy} &=& \int_{\epsilon=0}^1
d\epsilon \; \sum_{q,g} P_{q,g}(\epsilon)
\frac{1}{ (1 - (1-f_{q,g}) \cdot \epsilon)^2} \nonumber \\
&&\times
\frac{\sigma^{NN}_{q,g}(R,\omega^{\min})} {d^2E^\prime_Tdy} \; ,
\label{eq:sigma-AA}
\end{eqnarray}
where $E^\prime_T = E_T/(1 - (1-f_{q,g})\cdot \epsilon)$. The
$(1-f_{q,g}) \cdot \epsilon$  factor accounts for the total
''missed'' energy in a jet cone measurement, which
necessitates $E^\prime_T > E_T$, and the Jacobian $J= |d^2E^\prime_T/d^2E_T|$ is
properly accounted for. In this paper possible fluctuations of $f_{q,g}$ independent
of $\epsilon$  are not considered. Simple analytic limits illustrate the physics
represented by Eq.~(\ref{eq:sigma-AA}): if there is no energy loss,
$P(\epsilon) =  \delta(\epsilon)$, $J=1$ and the cross section is unaltered. In the
opposite limit, $P(\epsilon) =  \delta(\epsilon-1)$, the quenching of jets, if any,
is completely determined by the fraction of the lost energy $f_{q,g}$ that is recovered
in the experimental acceptance. When  $f_{q,g}=1$ once again there will be no
attenuation of jets and when  $f_{q,g} \rightarrow 0$ our result approximates the
inclusive particle $R_{AA}(p_T)$~\cite{Vitev:2005he}, see also Eq.~(\ref{Rjet}).

Next, we obtain the full jet shape, including the contributions
from the vacuum and the medium-induced bremsstrahlung:
\begin{eqnarray}
\psi_{\rm tot.}\left({r}/{R}\right) &=&
\frac{1}{\rm Norm}  \int_{\epsilon=0}^1
d\epsilon \; \sum_{q,g} P_{q,g}(\epsilon)
\frac{1}{ (1 - (1-f_{q,g}) \cdot \epsilon)^3} \nonumber \\
&&\times
\frac{\sigma^{NN}_{q,g}(R,\omega^{\min})} {d^2E^\prime_Tdy}
\Big[ (1- \epsilon) \;
\psi_{\rm vac.}^{q,g}\left({r}/{R}\right)
\nonumber \\
&& \hspace*{2.7cm}  +  \, f_{q,g}\cdot \epsilon \;
\psi_{\rm med.}^{q,g}\left(r/R\right) \Big] \; . \qquad
\label{psitotmed}
\end{eqnarray}
We recall that, by definition, the area under any  differential
jet shape, $\psi_{\rm tot.}\left({r}/{R}\right)$,
$\psi_{\rm vac.}\left({r}/{R}\right)$ and $\psi_{\rm med.}\left({r}/{R}\right)$,
is normalized to unity. Integrating over $r$ in Eq.~(\ref{psitotmed}),
it is easy to see that the correct ``Norm'' is the quenched
cross section,  Eq.~(\ref{eq:sigma-AA}). The interested reader can
independently carry out the analysis of the simple limiting cases
and gain insight into the dominant contribution to the full
jet shape. Proper treatment of isospin is implicit in
Eqs.~(\ref{eq:sigma-AA})  and~(\ref{psitotmed}).

\subsection{Energy sum rule}

Sum rules provide useful integral representation of conservations
laws, originating from symmetries in QCD. One such example is
momentum conservation in independent fragmentation:
\begin{eqnarray}
&& \sum_h \int_0^1 z D_{h/q,g}(z,Q^2)\,dz = 1 \; ,
\label{SR:FF-1}
\end{eqnarray}
where $z=p_{T,h}/p_{T,q(g)}$ is the momentum  fraction of parent
partons carried by fragmentation hadrons. The same sum rule
will hold in the presence of a medium since the total momentum
of the partons must be conserved irrespective of whether  they
are propagating in vacuum or in the QGP with/without medium-induced
bremsstrahlung~\cite{Vitev:2005he,Vitev:2005yg}. For jets, taking
a monochromatic pulse in the vacuum,
\begin{eqnarray}
\frac{1}{\sigma} \frac{\sigma^{NN}}{d^2E_T} = \delta^2(\bf{E_T}
- \bf{E_0}) \; ,
\label{SR:CS-pp}
\end{eqnarray}
we can easily verify that in the presence of a QGP
\begin{eqnarray}
\int d^2 E_T\frac{1}{\sigma}\frac{\sigma^{AA}
(R\rightarrow \infty, \omega^{\min} \rightarrow 0)}{d^2E_T} E_T
= E_0 \; ,
\label{SR:CS-pp}
\end{eqnarray}
in case of perfect experimental acceptance. More generally,
only a fraction, $1-(1-f) \langle \epsilon \rangle$, of $E_0$
is recovered.


\subsection{Numerical results}

\begin{figure}[!t]
\centerline{\hbox{
\epsfig{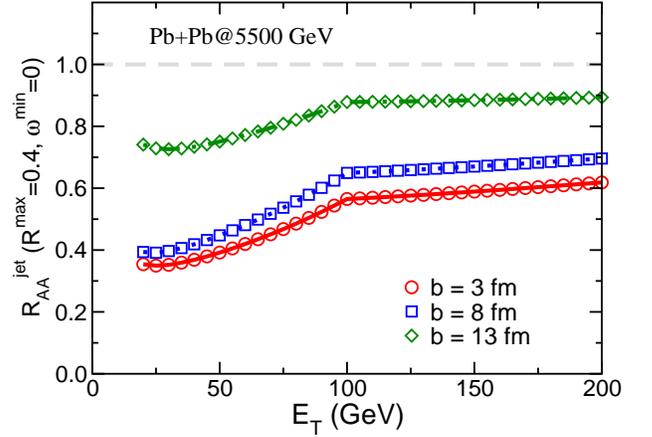}
}}
\caption{(Color online) Nuclear modification factor
$R_{AA}^{\text{jet}}(R^{\max},\omega^{\min})$ as a function of the jet
transverse  energy, $E_T$, at impact parameters $b=3$~fm (circle), $b=8$~fm
(square) and $b=13$~fm (diamond) in Pb+Pb collisions with $\sqrt{s}=5.5$~TeV.
}
\label{fig:Raa-1}
\end{figure}

Combining  our  full theoretical model for the jet shape and the jet
cross section in heavy  ion  collisions  with realistic numerical
simulations of parton propagation in the  QGP, see Section
\ref{subsec:QGP}, we first evaluate the  nuclear modification factor
$R_{AA}^{\text{jet}}(R^{\max},\omega^{\min})$ in Pb+Pb collisions
with center of mass energy $\sqrt{s}=5.5$~TeV at the LHC.
Fig.~\ref{fig:Raa-1} illustrates the attenuation of the measured jet
rate as a function of the jet energy $E_T$ for different centrality
classes. We use impact parameters $b = 3, 8, 13$~fm in conjunction
with a jet cone radius $R^{\max} = 0.4$  and no acceptance cut
($\omega^{\min}=0$~GeV). The evolution of
$R_{AA}^{\text{jet}}(R^{\max},\omega^{\min})$  for jets is similar
to the one for leading particles in that $\ln R_{AA} \approx -
\kappa N_{part.}^{2/3}$~\cite{Vitev:2005he}. However, $\kappa$ will
depend on the selection of  $R^{\max}$ and $\omega^{\min}$ in
addition to the steepness of the underlying jet spectra and the
properties of the  QGP. We recall that the energy of the jet that
will be redistributed  out of the cone is $\epsilon (1-f)$, see Eq.
(\ref{eq:sigma-AA}), and the variation of its quenching with
centrality is related to the fractional parton energy loss
$\epsilon=\Delta E/E$ and its fluctuations, given by $P(\epsilon)$.

\begin{figure}[!t]
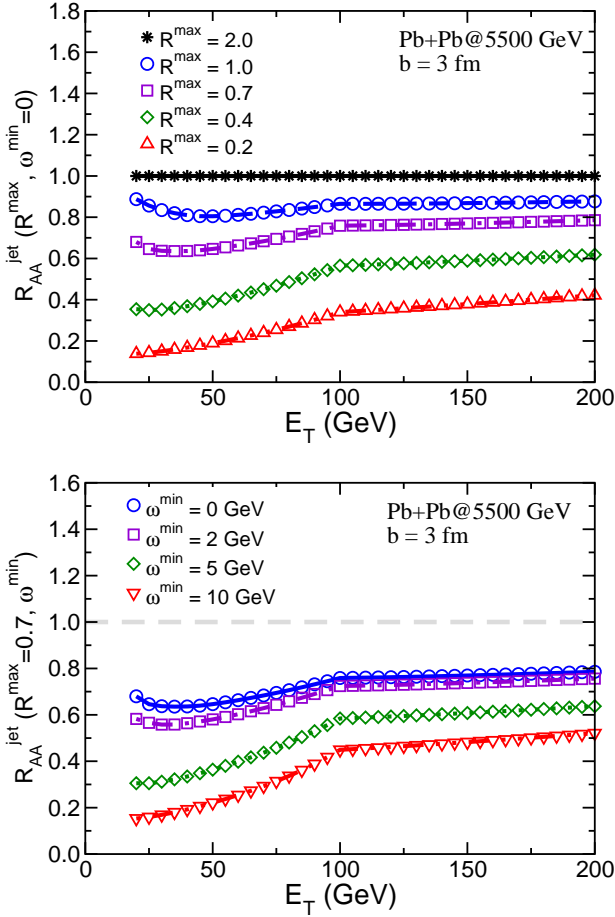

\centerline{\hbox{
$\begin{array}{c}
\\[2ex]
\epsfig{file=Raa-R-b3-jet.eps,width=3.2in,clip=,angle=0} \\[2ex]
\epsfig{file=Raa-Wmin-b3-jet.eps,width=3.2in,clip=,angle=0} \\[2ex]
\end{array}$
}}
\caption{(Color online) $E_T$-dependent nuclear modification factor
$R_{AA}^{\text{jet}}(R^{\max},\omega^{\min})$ for different jet cone radii
$R^{\max}$ (top panel) and at different acceptance cuts $\omega^{\min}$
(bottom panel) in $b=3$~fm  Pb+Pb collisions at $\sqrt{s}=5.5$~TeV.
}
\label{fig:Raa-2}
\end{figure}

Fig.~\ref{fig:Raa-2} demonstrates the sensitivity of
$R_{AA}^{\text{jet}}(R^{\max},\omega^{\min})$ to the properties of the medium-induced
gluon radiation through the independent variation of  $\omega^{\min}$
and  $R^{\max}$, advocated in this paper. For a fixed impact parameter,
$b=3$~fm, the top panel shows a study of the quenching strength versus
the jet cone radius when $\omega^{\min}=0$~GeV. In the approximations that we
employ $R^{\max} = 2$ is the upper bound of the medium-induced bremsstrahlung
opening angle relative to the jet and, consequently, constitutes perfect
experimental acceptance. In this case there is no deviation from
binary collisions scaling. The smooth evolution of $R_{AA}^{\text{jet}}(R^{\max},\omega^{\min})$
with decreasing $R^{\max}$ is a signature of the large-angle gluon radiation
pattern in the QGP~\cite{GLV,Vitev:2005yg}. Note that if $dI^g/d\omega dr$ were
predominantly collinear, there would be no deviation from unity.
For $ R^{\max} \leq 0.2 $  the magnitude of jet  quenching approaches the
suppression for leading hadrons. A good starting point is a cone radius
selection $R^{\max} = 0.4 - 0.7$ if the experimental statistics allows for
positive identification of 30\% to a factor of 2 variation in the measured cross
section. In the bottom panel of Fig.~\ref{fig:Raa-2} we present the
sensitivity of jet attenuation to the minimum particle momentum/calorimeter
tower energy deposition cut $\omega^{\min}$. For a finite $R^{\max} = 0.7$ even
if $\omega^{\min} = 0$~GeV  $R_{AA}^{\text{jet}}(R^{\max},\omega^{\min})$ does not
reach unity, see our discussion above. The largest variation in the quenching
strength is observed between $\omega^{\min} = 2$~GeV and  $\omega^{\min} =
 5 - 10$~GeV, and reflects the typical energy of the stimulated gluon
emissions. We emphasize that, in the GLV approach~\cite{GLV}, partons
lose energy through $\sim$~few GeV bremsstrahlung gluons~\cite{Vitev:2005he}.
For  $\omega^{\min} > 10$~GeV, $R_{AA}^{\text{jet}}(R^{\max},\omega^{\min})$
approaches again the characteristic leading particle suppression.
In summary, for the same centrality, $E_T$  and $\sqrt{s}$ the continuous
variation of quenching values may  help differentiate between competing
models of parton energy loss~\cite{Majumder:2007iu}, thereby eliminating
the order of magnitude uncertainty in the extraction of the QGP density.

Detailed investigation of $R_{AA}^{\text{jet}}(R^{\max},\omega^{\min})$  can
also indicate whether ``elastic'' $2 \rightarrow 2$ processes, such as
collisional energy loss~\cite{Wicks:2007zz}, or ``inelastic'' $2 \rightarrow 2+n $
processes, such as bremsstrahlung~\cite{Gyulassy:2003mc} and hadron
dissociation in the QGP~\cite{Adil:2006ra}, dominate the inclusive particle
and particle correlation quenching. If the energy loss {\em per interaction}
in the first scenario $\Delta E_{\rm coll.} / E \leq 5\%$, the recoil parton form
the medium will accelerate almost transversely relative to the jet axis and will not be
part of the jet for any reasonable selection of $R^{\max}$. Therefore, for
collisional energy loss, in contrast to the well-defined evolution of the
jet suppression with cone radius and the acceptance cut seen in Fig.~\ref{fig:Raa-2},
the cross section attenuation will be large and constant and will approximate
the quenching of leading hadrons. Note that $R_{AA}^{\rm jet} < 1$ has also
been observed in Monte-Carlo silumatios of jet quenching~\cite{Lokhtin:2007ga}.

\begin{figure}[!ht]
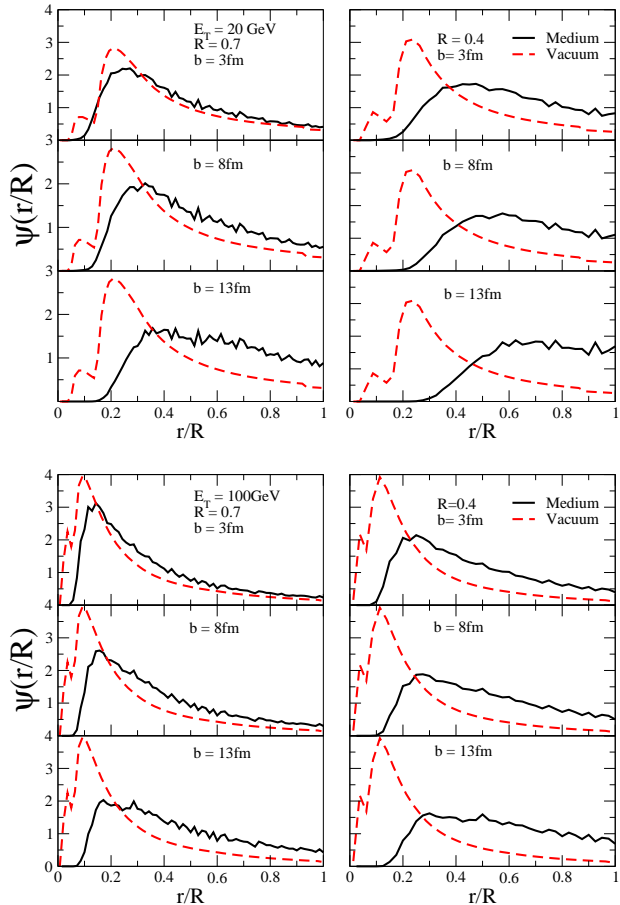

\centerline{\hbox{
$\begin{array}{c}
\\[2ex]
\epsfig{file=Psi-b-e20.eps,width=3.2in,clip=,angle=0} \\[2ex]
\epsfig{file=Psi-b-e100.eps,width=3.2in,clip=,angle=0} \\[2ex]
\end{array}$
}}
\caption{(Color online) The vacuum and medium-induced only
jet shape for $E_T =  20$~GeV (top panel)  and $E_T = 100$~GeV (bottom panel)
at impact parameters $b=3,\, 8,\, 13 $~fm in Pb+Pb collisions at the LHC.
Two jet cone radii, R = 0.7 and 0.4 are shown.
}
\label{fig:Rpsi-medium}
\end{figure}

We now turn to the numerical results for the jet shape in Pb+Pb  collisions
at $\sqrt{s}=5.5$~TeV at the LHC. In Fig.~\ref{fig:Rpsi-medium} we
first explore the difference between the vacuum and the medium-induced
only ($E_T$ given for the parent parton) 
$\psi(r/R)$ as a function of the impact parameter, jet energy, and
the cone radius. We note that in central heavy ion reactions for lower $E_T$
and, in particular, for $R \geq 0.7$ the two differential shapes can be quite
similar. The differences become more pronounced for smaller jet radii
where the experimental acceptance will subtend the part of phase space
with the most effective cancellation of the collinear medium-induced
radiation~\cite{Vitev:2005yg}. It is interesting to observe that in going
to more peripheral collisions $\psi_{\rm med.}(r/R)$
becomes slightly wider. The underlying reason is that the LPM destructive
interference between the radiation induced by the large $Q^2$ scattering and
the radiation induced by the subsequent interactions in the QGP determines
the angular distribution in the bremsstrahlung spectrum.  Thus, a  small
medium size facilitates the resulting cancellation for gluons of large
formation time. A general observation is that the medium tends to redistribute
the flow of energy more evenly inside the jet cone, especially for large
$r/R \rightarrow 1 $

\begin{figure}[!ht]
\centerline{\hbox{
\epsfig{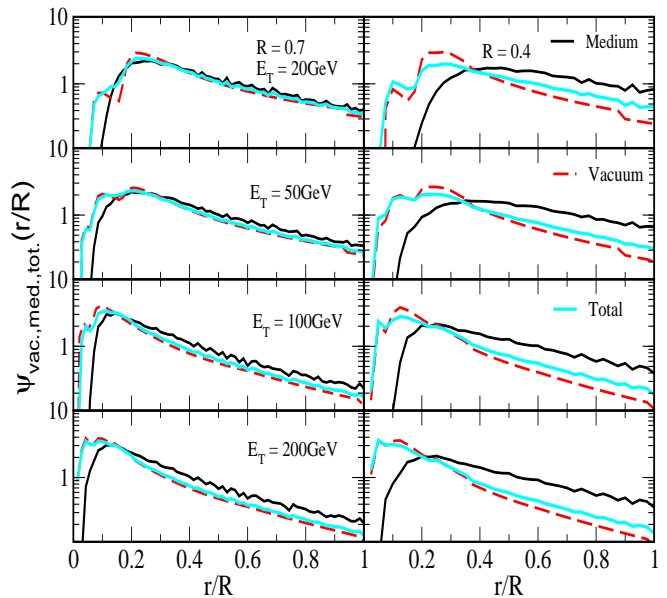}
}}
\caption{(Color online) Comparisons of the jet shape in vacuum,
the medium-induced jet shape, and the total jet shape for cone radii
$R=0.7$ and $R=0.4$ and four different  energies
$E_T=20, \, 50, \, 100, \, 200$~GeV, respectively, in central
Pb+Pb collisions at the LHC.}
\label{fig:Rpsi-jet}
\end{figure}

The pattern of energy flow for in-medium jets is shown in Fig.~\ref{fig:Rpsi-jet}
together with $\psi_{\rm vac.}(r/R)$ and $\psi_{\rm med.}(r/R)$ for comparison.
We used  $E_T = 20$~GeV  to $200$~GeV and $R = 0.4$ to $0.7$ to cover a wide
range of measurements that will become accessible during the first year of
heavy ion running at the LHC. One observes that there is no significant
distinction  between the jet shape in the vacuum and the total in-medium $\psi(r/R)$.
The underlining reason for this surprising result is that although
medium-induced gluon radiation produces a broader $\psi_{\rm med.}(r/R)$,
this effect is offset by the fact that the jets lose a finite
amount of their energy, see Figs.~\ref{fig:ERatio-3d} and~\ref{fig:ERatio-2d}.
Furthermore, when part of the lost energy is missed due to finite experimental
acceptance, the required higher initial virtuality jets are inherently
narrower, see Figs.~\ref{fig:LHCshapes} and~\ref{fig:LHCshapes-3D}.

\begin{table*}[!tb]
\begin{center}
\begin{tabular}{||c||c|c|c||}
\hline
   $ R=0.4 $   &  Vacuum  &  Complete E-loss  & Realistic case   \\
\hline
   $\langle r/R \rangle$, $E_T=20$GeV   &  0.41   & 0.55   &  0.45   \\
\hline
   $\langle r/R \rangle$, $E_T=50$GeV   &  0.35   & 0.48   &  0.38  \\
\hline
   $\langle r/R \rangle$, $E_T=100$GeV  &  0.28   & 0.44   &  0.32  \\
\hline
   $\langle r/R \rangle$, $E_T=200$GeV  &  0.25   & 0.40   &  0.28  \\
\hline \hline
   $ R=0.7 $   &  Vacuum &  Complete E-loss  & Realistic case   \\
\hline
   $\langle r/R \rangle$, $E_T=20$GeV   &  0.41    & 0.44   &  0.42 \\
\hline
   $\langle r/R \rangle$, $E_T=50$GeV   &  0.33   & 0.39   &  0.37  \\
\hline
   $\langle r/R \rangle$, $E_T=100$GeV  &  0.27   & 0.34   &  0.29  \\
\hline
   $\langle r/R \rangle$, $E_T=200$GeV  &  0.24   & 0.31   &  0.26  \\
\hline
\end{tabular}
\end{center}
\caption{ Summary of mean relative jet radii $\langle r/R \rangle$
in the vacuum, with complete energy loss, and  in the QGP medium.
Shown are results for cone radii $R=0.4$ and $R=0.7$ and  transverse
energies $E_T = 20, 50, 100, 200$~GeV at $ \sqrt{s}=5.5$~TeV central
Pb+Pb collisions at the LHC. } \label{table:mean-radii}
\end{table*}

In Table~\ref{table:mean-radii} we show the mean relative jet
radii~$ \langle r/R \rangle $ in the vacuum and in the QGP medium
created at the LHC for  two different cone selections $R=0.4$ and
$R=0.7$ and four transverse energies  $E_T = 20,\, 50,\,  100,\,
200$~GeV. We see that in the realistic numerical simulation there is
very little $<10\%$ increase in the magnitude of this observable.
The difference is slightly larger for a smaller cone, since it
emphasizes the large-angle character of the medium-induced
radiation~\cite{GLV,Vitev:2005yg}. Therefore, a rough 1-parameter
characterization of energy flow in jets will not resolve the effect
of the QGP medium. It can, however, exclude simplistic scenarios of
full jet stopping in the QGP that lead to $\langle r/R \rangle$ growth by as much
as 60\%.  It is also important to stress that the QGP is rather
``gray'' than ``black'' and only a fraction of the energy of the
parent paron is lost via stimulated gluon emission. The effect of
even a moderate $\Delta E^{in}(R^{\max},\omega^{\min})/E$ can be
amplified by the steeply falling cross sections for the
$R_{AA}^{\texttt{jet}}(E_T; R^{\max},\omega^{\min})$ observable, see
Fig.~\ref{fig:Raa-2},  but this is not the case for $\langle r/R \rangle$.

\begin{figure}[!b]
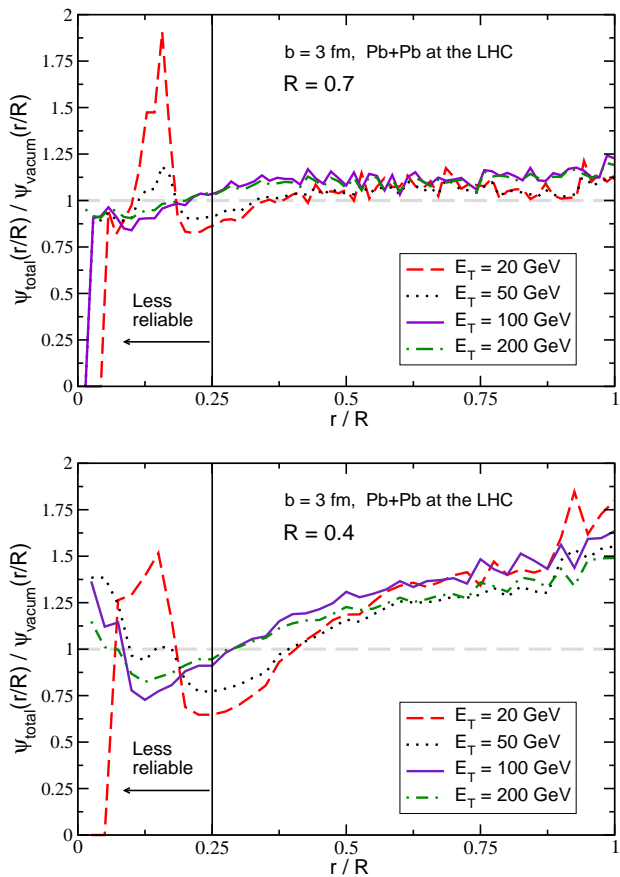

\centerline{\hbox{
$\begin{array}{c}
\\[2ex]
\epsfig{file=Rpsi-R0.7-W0-jet-N.eps,height=2.2in,width=3.2in,clip=,angle=0} \\[2ex]
\epsfig{file=Rpsi-R0.4-W0-jet-N.eps,height=2.2in,width=3.2in,clip=,angle=0} \\[2ex]
\end{array}$
}}
\caption{(Color online) The ratios of total jet shape in heavy-ion
collisions to the jet shape in the vacuum for  jet energies
$E_T=20,\, 50,\, 100,\, 200$~GeV. Two cone radii $R=0.7$ (top panel) and
$R=0.4$ (bottom panel) at $b=3$~fm in Pb+Pb  collision with
$\sqrt{s}=5.5$~TeV were chosen.
}
\label{fig:Rpsi-ratio}
\end{figure}

Lastly, we point out where the anticipated jet broadening effects will be
observed in the differential shape by studying  the ratio
$\psi_{\rm tot.}(r/R)/\psi_{\rm vac.}(r/R) $ in Fig.~\ref{fig:Rpsi-ratio}.
We have used the same transverse energies and cone radii as in Fig.~\ref{fig:Rpsi-jet}.
We recall that the small $r/R < 0.25$ region of the intra-jet energy flow
in p+p collisions  in our calculation has uncertainties associated
with the normalization of the jet shape. In the moderate and large $r/R>0.25$
region our theoretical  model gives excellent descriptions of the
Fermilab  Run II (CDF II) data, as shown in Fig.~\ref{fig:CDF}.
The QGP effects are manifest in  the
``tails'' of the energy flow distribution and for a cone radius $R=0.4$
the ratio could reaches  $\sim 1.75$ when $r/R \rightarrow 1$. However, for
experiments to observe this enhancement of the ratio of the total jet shape
in medium to jet shape in vacuum at $ r/R > 0.5$,  high statistics
measurements will be needed. This precision can hopefully
be achieved with the large acceptance experiments at the LHC.

\section{Conclusions}
\label{sec:summary}

The unprecedentedly high center of mass energies at the LHC will usher in a new 
era of precision many-body QCD. The theory and phenomenology of jets in nuclear
collisions are expected to evolve as the new frontier in the perturbative studies
of parton propagation in the QGP~\cite{Vitev:2008jh}. In this paper we discussed 
three important  aspects of such studies: a generalization of the analytic approach 
for calculating differential jet shapes~\cite{Seymour:1997kj} that can accommodate 
experimental acceptance cuts needed to isolate jets in the high multiplicity 
environment of heavy ion collisions; the theory of the intra-jet energy flow 
redistribution through large-angle medium-induced gluon bremsstrahlung; and 
a comprehensive new set of experimental observables that can help identify and 
characterize the mechanisms of parton interaction in nuclear matter.

In elementary nucleon-nucleon collisions we compared our theoretical model to
the CDF II Tevatron data on jet shapes~\cite{Acosta:2005ix} and investigated the 
baseline $\psi_{\rm vac.}(r/R)$ at the higher $\sqrt{s}=5.5$~TeV at the LHC. We 
found that in the absence of a hot and dense QGP matter these shapes are self-similar 
and approximately independent of the cone radius $R$. Elimination of
low momentum particles of up to $\sim$few GeV is not likely to significantly alter
the pattern of intra-jet energy flow for $E_T > 50$~GeV jets. In nucleus-nucleus 
reactions we demonstrated that the characteristic large-angle QGP-stimulated
gluon emission~\cite{Vitev:2005yg} persists to all orders in the correlation between 
the elementary bremsstrahlung sources. We showed  that this intensity spectrum 
can be fully characterized by the amount of the lost energy that falls inside 
the jet cone ($ r < R^{\max}, \, \omega > \omega^{\min}$) and derived the medium modification of the
jet shapes and jet cross sections in the QGP, subject to an intuitive energy sum rule.

To demonstrate the connection between the QGP properties, the mechanisms of parton 
interaction and energy loss in hot and dense matter, and a new class of jet-related 
experimental observables, we carried out realistic simulations of quark and gluon 
production and propagation in the medium created in relativistic heavy ion 
collisions at the LHC. We introduced a natural generalization of the leading particle 
suppression to jets and showed that it is a more differential and powerful tool 
that can be used to assess  in approximately 
model-independent way the characteristic properties of the induced 
gluon intensify spectrum. Consequently, in the future  progress can be made toward 
identifying the set of approximations~\cite{Gyulassy:2003mc,BDMPS-Z-ASW,GLV,HT} 
that most adequately reflect the dynamics of hard probes in the QGP.
We also discussed how the evolution of $R_{AA}^{\texttt{jet}}(R^{\max},\omega^{\min})$
with the jet cone radius $R^{\max}$ and the acceptance cut $\omega^{\min}$, or the 
lack thereof, can help differentiate between radiative and collisional energy loss 
paradigms of light and heavy quark attenuation. The theoretical approach, developed
in this manuscript, allows to investigate the correlation between the quenching 
of jets and the in-medium modification of their shape. Surprisingly, up to five-fold 
attenuation of the cross section corresponds to a rather modest  $\leq 10\%$  growth 
of the mean relative cone radius $\langle r/R \rangle$. The anticipated broadening
of jets is most readily manifest  in the periphery of the cone, $r/R \rightarrow 1$,  
and for smaller radii, e.g.  $R^{\max}=0.4$.

Further refinements in jet phenomenology, especially the consideration of jet cross 
sections, should include cold nuclear matter effects, such as nuclear shadowing, 
the  Cronin effect, and initial state energy loss~\cite{Vitev:2008vk}. 
We finally note that the study of inclusive jet shapes and cross sections in heavy
ion collisions can easily be generalized to hadron, photon or di-lepton tagged
jets~\cite{D'Enterria:2007xr,Mironov:2007bu,Grau:2008ed}  with the benefit of    
additional constraints on the hard process virtuality and the parton energy.


\begin{acknowledgments}
We thank M. H. Seymour, H. Caines, N.~Grau and H.~Takai for many helpful 
discussions. This research is  supported by the US Department of Energy, Office 
of Science, under Contract No. DE-AC52-06NA25396 and in part by the LDRD program 
at LANL, the NNSF of China and the MOE of China under Project No. IRT0624.

\end{acknowledgments}

\begin{appendix}

\section{Jet cross sections}
\label{app:cross}

In this paper we focus exclusively on large momentum transfer
processes, $Q^2 \gg \Lambda_{\rm QCD}^2$, that can be
systematically calculated in the framework of a reliable theory,
the perturbative QCD factorization approach. Factorization
not only separates the short- and long-distance QCD dynamics but
implies universality of the parton distribution functions (PDFs)
and fragmentation functions (FFs) and infrared safety of the
hard scattering cross sections. For hadronic collisions,
one of the most inclusive processes is jet production. To lowest
order (LO) the invariant differential cross section
reads~\cite{Vitev:2006bi}:
\begin{eqnarray}
\frac{ d\sigma_{ h_a h_b } }{ dy_c d^2 p_{T_c} }  &=& K \sum_{abcd}
\int\limits_{y_{d\,\min}}^{y_{d\,\max}}
dy_d \, \frac{\phi_{a/h_a}({x}_a,\mu_f) \phi_{b/h_b} ({x}_b,\mu_f) }
{{x}_a{x}_b}   \, \nonumber \\
&& \times \frac{\alpha_s^2(\mu_r)}{{s}^2 }
|\overline {M}_{ab\rightarrow cd}|^2   \;.
\label{basicjet}
\end{eqnarray}
Here,  $s = (P_a + P_b)^2$  is the squared center of mass
energy of the hadronic collision  and  $x_a = p_a^+/P_a^+$,
$x_b = p_b^-/P_b^-$  are the lightcone momentum fractions of the
incoming partons. In this formulation, for massless initial-state
quarks and gluons,
\begin{equation}
y_{d\, \max (\min) } = + (-) \ln \left( \frac{\sqrt{s}}{m_{T\,d}}
- \frac{m_{T\,c}}{m_{T\,d}} e^{+(-)y_c} \right) \; ,
\label{limits}
\end{equation}
where $m_{T_i}^2 = m_i^2 + p_{T_i}^2$. In Eq.~(\ref{basicjet})
$\phi_{i/h_i}(x_i, \mu_{f\, i})$ is the distribution function
of parton ``$i$'' in the hadron  $h_i$ and  $\mu_{r}$ and $\mu_{f\, i}$
are the  renormalization and factorization and scales, respectively.
In this work calculations are done strictly in the collinear
factorization approach and  we use the  CTEQ6.1 LO
PDFs~\cite{Pumplin:2002vw}. $ |\overline {M}_{ab\rightarrow cd}|^2 $
are the squared  matrix elements for $ab \rightarrow cd $
partonic sub-processes.

\begin{figure}[!t]
\epsfig{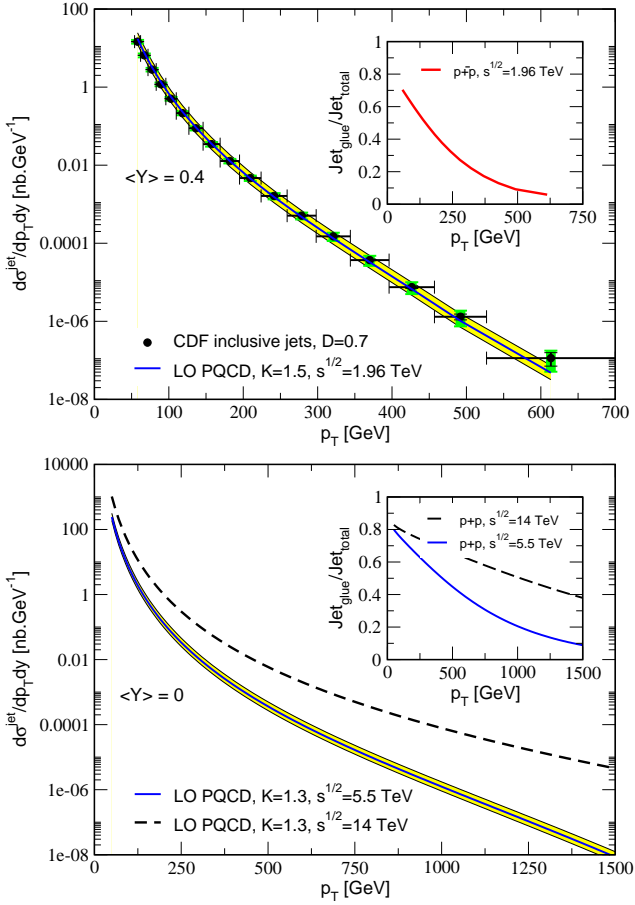}
\caption{ Top panel: Inclusive jet cross section in p+$\bar{\rm p}$
collisions
at the Tevatron $\sqrt{s} = 1.96$~GeV calculated to LO in PQCD and
compared to the CDF run II data~\cite{Abulencia:2005jw}. Insert shows
the fraction of gluon jets as a function of $p_T$. Bottom panel:
predicted
baseline jet cross sections in p+p collisions at the LHC at
$\sqrt{s} = 5.5$~TeV and 14 TeV.
}
\label{sigmajet1}
\end{figure}

Numerical results for inclusive jet cross sections in high
energy hadronic collisions are shown in Fig.~\ref{sigmajet1},
here $p_T = E_T$.
The top panel compares the LO calculation, Eq.~(\ref{basicjet}),
to CDF data on jet cross sections in ${\rm p}+\bar{\rm p}$ at
$\sqrt{s}=1.96$~TeV. Excellent agreement between data and the
theory using $K=1.5$, independently extracted from the charged
hadron $h^+ + h^-$ differential cross section at the Tevatron.
It indicates $ \sim 50\% $ next-to-leading order correction.
Alternatively,
we have studied the sensitivity of the cross section to the
choice of the factorization and renormalization scales by
varying $\mu_r = \mu_f = p_T/2$, $p_T$ and
$2 p_T$. Not surprisingly, the uncertainty is also on the order of
$ \sim 50\%$, similar to the phenomenological $K$-factor.
Jet shapes depend on the parton species, quarks versus gluons,
and the insert shows the fraction of quark jets versus $p_T$
at the Tevatron. The bottom panel gives predictions for the
corresponding jet cross sections at the LHC per nucleon-nucleon
collision for $\sqrt{s} = 14$~TeV and 5.5~TeV, without quenching.
Insert shows the increased fraction of gluon jets relative to the
Tevatron.

We can now evaluate the feasibility of differential jet shape
measurements
at the LHC. During the first three years of running, even at a
fraction of the
designed  ${\cal L} = 10^{34}$~cm$^{-2}$s$^{-1}$, LHC is expected to
deliver
an integrated luminosity of 10 fb$^{-1}$ per year. In heavy ion
collisions,
nominal ${\cal L} = 10^{27}$~cm$^{-2}$s$^{-1}$ is not expected to be
achieved
either.  An integrated luminosity of 1 nb$^{-1}$ per year is  a
realistic
projection.  As shown in this paper, the anticipated quenching
factor  for
energetic jets  depends on the selection of $R^{\max}$ and $
\omega^{\min}$.
In the limit of narrow jets  $R_{AA}^{jet} \approx R_{AA}^{h^\pm}
   = 0.25 - 0.5$~\cite{Vitev:2005he,Wicks:2007zz}.
Taking this into account, but neglecting for the  moment the
complications
associated with jet reconstruction in the high particle multiplicity
environment of heavy ion collisions~\cite{Salur:2008hs}, from Fig.~
\ref{sigmajet1}
we find that excellent $ < 10 \% $ statistical precision can be
achieved for
inclusive measurements of jets of $p_T$ as high as 160 GeV in Pb+Pb
reactions
and 1.3 TeV in p+p reactions. Jet {\em shape} measurements require
higher
statistics since the shape functions are precipitously falling as
of $r/R \rightarrow 1 $.  We expect that very good, $ < 30 \% $ at large
$ r/R \sim 1 $, {\em jet shape} measurements will be possible to $p_T$
as high
as 100 GeV and 900 GeV in Pb+Pb and p+p collisions, respectively.
These will
give an indication as to whether  medium-induced  broadening is present
in the tails of the intra-jet energy distribution. The estimates
presented
in this appendix are conservative, $\Delta y = 1$.

\section{Contributions to the vacuum jet shape}
\label{app:params}

\begin{figure}[!t] 

\centerline{\hbox{
\epsfig{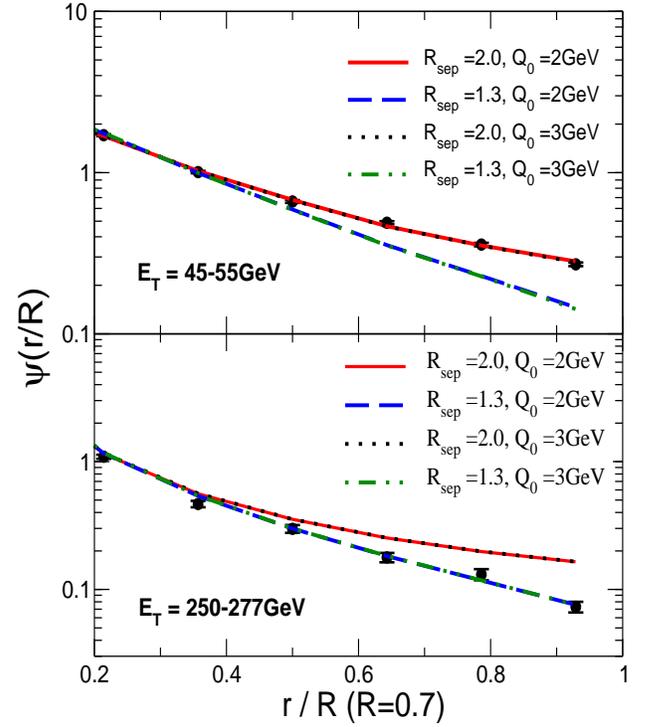}
}}
\caption{(Color online) Comparison of the theoretically computed
jet shapes with  $R_{sep}=1.3,\, 2$ and two
different non-perturbative correction scales $Q_0=2.0$~GeV
and $3.0$~GeV to experimental data~\cite{Acosta:2005ix}.
A jet cone radius  $R=0.7$ was chosen in $\sqrt{s}=1960$~GeV
    $p+{\bar p}$ collisions  by CDF II.}
\label{fig:Psi_Rsep_Q0}
\end{figure}

It is important that a jet finding algorithm be infrared and collinear
safe. In full Monte-Carlo simulations of high-energy hadronic events
the algorithm can be ``tested'' and matched to the experimental
measurement
techniques. In analytic calculations an approximate way to mimic the
effect of
jet splitting/merger is to  introduce an adjustable parameter, $R_{sep}
$,
for cone type algorithms. If two partons are within an angle $R_{sep}R
$ of
each other, they should be merged into one jet~\cite{Klasen:1997tj}.
This approach may not be optimal~\cite{Seymour:1997kj} since it does not
generalize intuitively for NLO jet shape calculations. When comparing
theoretical results to  experimental data  one finds
that $R_{sep}$ is a function of the event kinematics, i.e. it is  jet
momentum (energy) dependent. Nevertheless, at lowest order this is a
useful phenomenological approach to obtain the best possible description
of the baseline differential jet shapes in nucleon-nucleon collisions,
needed  for  the study of QGP-induced effects. Also, it known that
for jet cross sections at  NLO the results from other jet finding
algorithms,  such as the fully
infrared and collinear safe $k_T$ algorithm with a jet-size parameter
$D$,
coincide within a few \%  with the results from  the cone algorithm with
appropriate matching/choice of $R$ and $R_{sep}$~\cite{Ellis:2007ib}.

\begin{figure}[t!]
\centerline{\hbox{
\epsfig{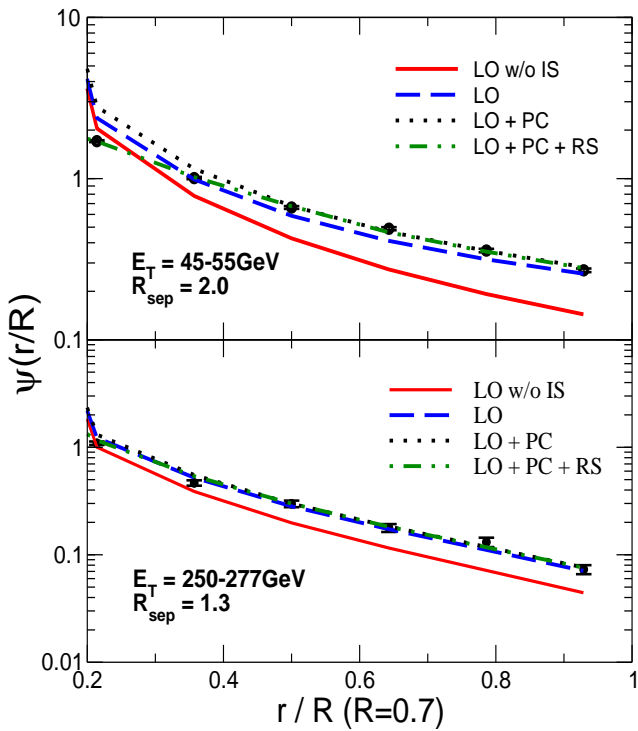}
}}
\label{plot:4}
\caption{(Color online) Differential jet shapes in $p+{\bar p}$
collisions at
$\sqrt{s}=1960$~GeV~\cite{Acosta:2005ix} are compared to theory.
Leading-order without
initial-state radiation (LO w/o IS), leading-order contibution (LO),
leading-order
   with power correction (LO+PC), and leading-order  with
power correction and Sudokov resummation (LO+PC+RS) results are shown
separately.}
\label{fig:Psi_LO+IS+PS+RS}
\end{figure}

In Fig.~\ref{fig:Psi_Rsep_Q0} we demonstrate how the numerical results
for jet shapes can be optimized  using experimental data at the
Tevatron.
Proper normalization of $\psi(r,R)$ in this Appendix is achieved via
first bin subtraction and does not affect the moderate and large $r/R$
part of the energy flow distribution. At low transverse energy,
$E_T=45-55$~GeV, a numerical calculation with large $R_{sep}=2.0$
gives a
much better fit than one with a small $R_{sep}=1.3$. At high
$E_T=250-277$~GeV, the theoretical result with small $R_{sep}=1.3$
agrees
with the data by CDF II fairly well. This finding is consistent with
a previous study showing that, to fit the data, at low jet $E_T$
one always needs a larger $R_{sep}$  and that $R_{sep}$ should
drop with increasing transverse energy~\cite{Klasen:1997tj}.
In the numerical calculations shown in Fig.~\ref{fig:Psi_Rsep_Q0}
the contribution of power correction to the jet shape has been included.
The scale $Q_0$ is used to separate the non-perturbative effect
from the perturbative derivation (see Eqs.~(\ref{eqn:alphas_0}) and
(\ref{eqn:Q0})). In Fig.~\ref{fig:Psi_Rsep_Q0} two different values
of the non-perturbative correction scale, $Q_0=2$~GeV and $Q_0=3$~GeV,
are used. It is clear that the curves with these two different scales
are practically indistinguishable, which demonstrates the consistency
of our treatment of non-perturbative effects.

In Fig.~\ref{fig:Psi_LO+IS+PS+RS} we illustrate the influence of
the different perturbative and non-perturbative contributions to
the jet shape for different $E_T$. We observe that the effect of
initial-state radiation (IS), absent in lepton colliders,  is sizable.
In fact,
at very high jet energy, a leading-order (LO) calculation with initial-
state
radiation already gives a good description of the experimental
data. However, below $\approx 75 $~GeV it is impossible to
fit the data using only leading order, even with the maximum
$R_{sep}=2$.
Other contributions should be considered in an improved theoretical
description of jet shapes. These include the effect of the running
coupling
constant in the momentum transfer integrals (MLLA) and
power corrections (PC) $\propto Q_0/E_T$.  Sudakov resummation (RS)
ensures the finiteness of $\psi(r,R)$ in the region $r \rightarrow 0$.
The explicit formulas for these two contributions are given in
Section~\ref{subsec:theory}. It can be seen from
Fig.~\ref{fig:Psi_LO+IS+PS+RS}
that all perturbative and non-perturbative effects should be taken
into account
if reliable description of the experimental results is to be achieved.


\vspace*{1cm}

\section{Double differential medium-induced jet shape}
\label{app:dz}

\begin{figure*}[!t]
\epsfig{file=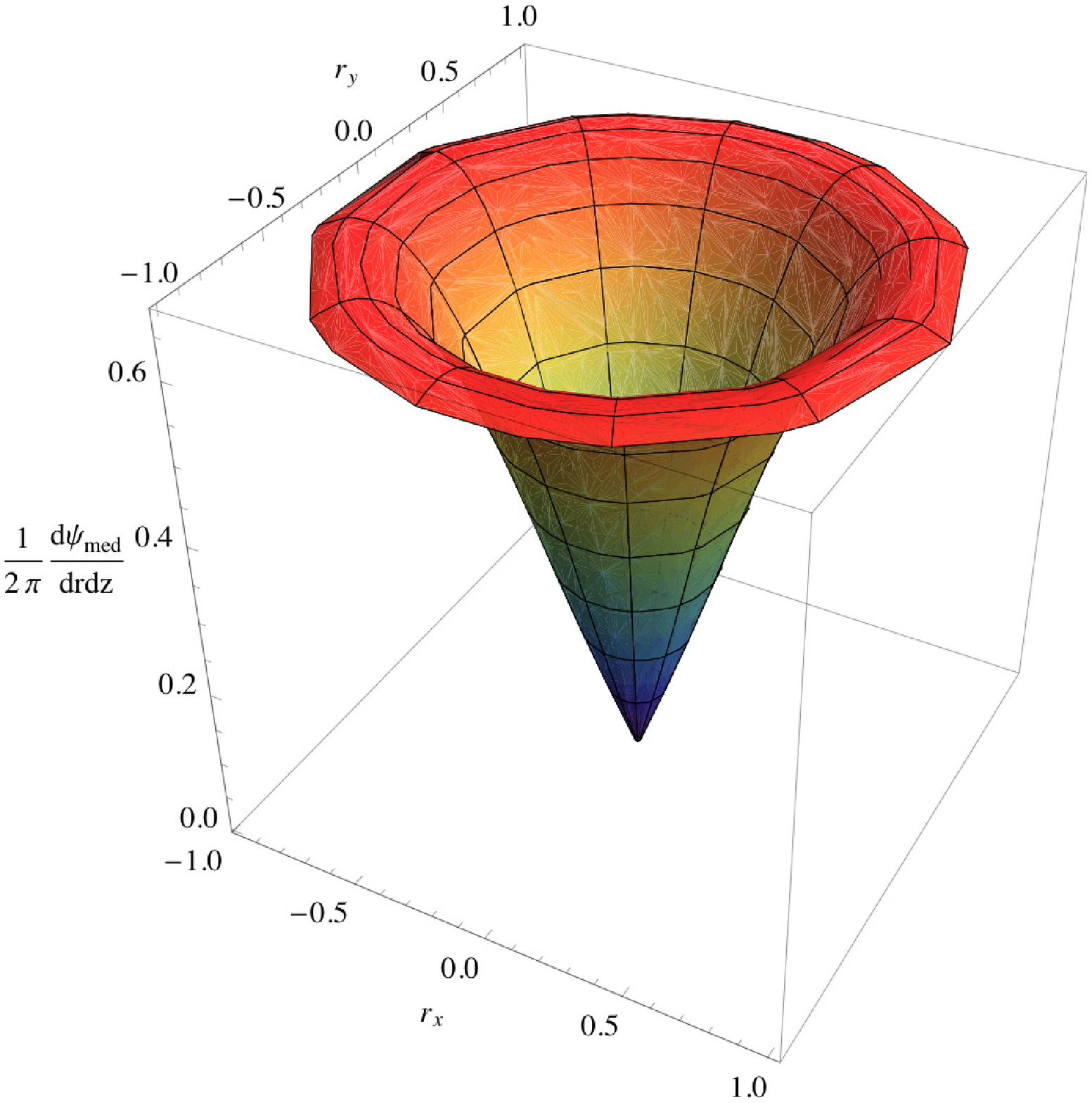,height=2.5in,clip=,angle=0}
\epsfig{file=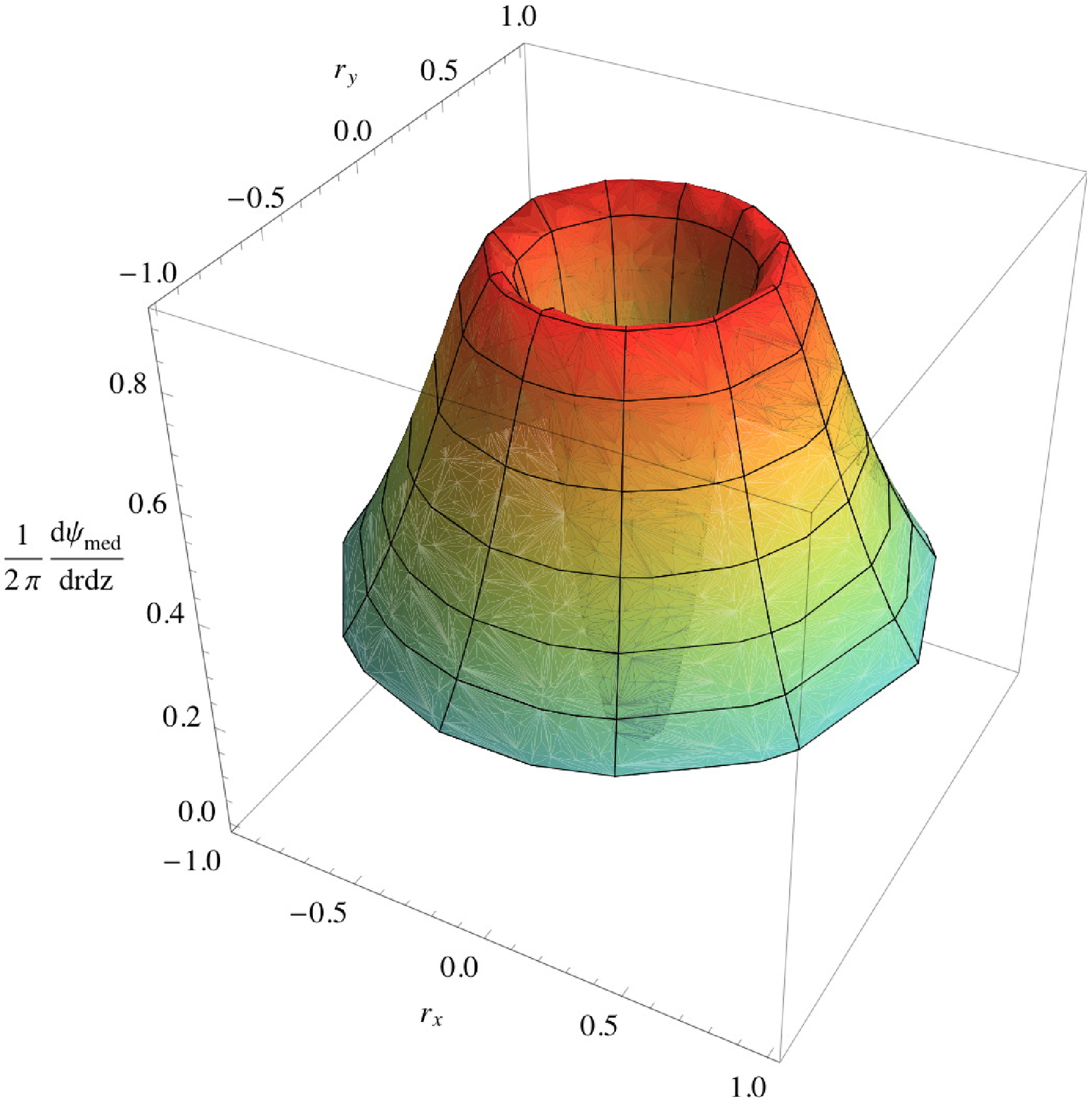,height=2.5in,clip=,angle=0}\\
\epsfig{file=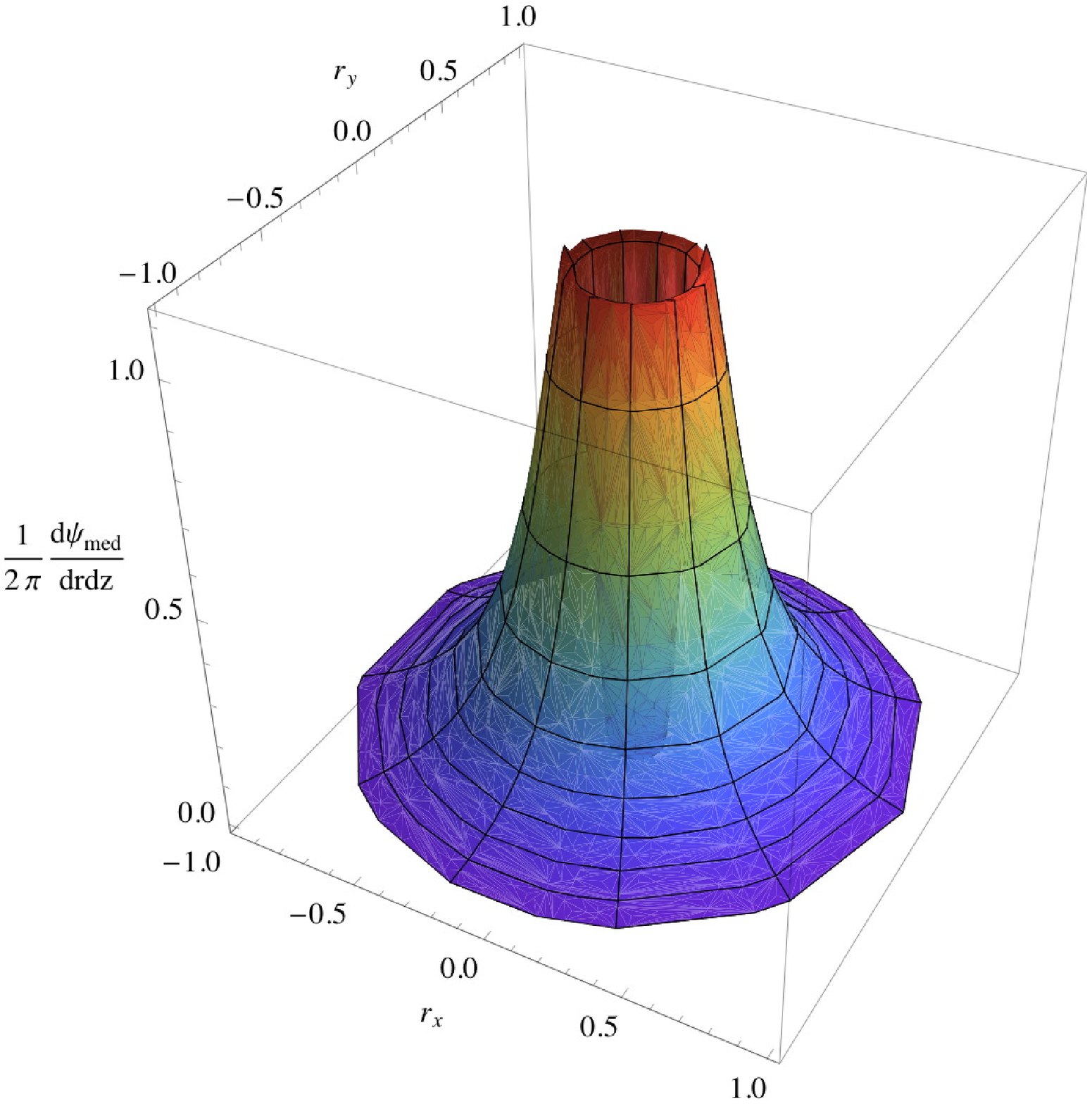,height=2.5in,clip=,angle=0}
\epsfig{file=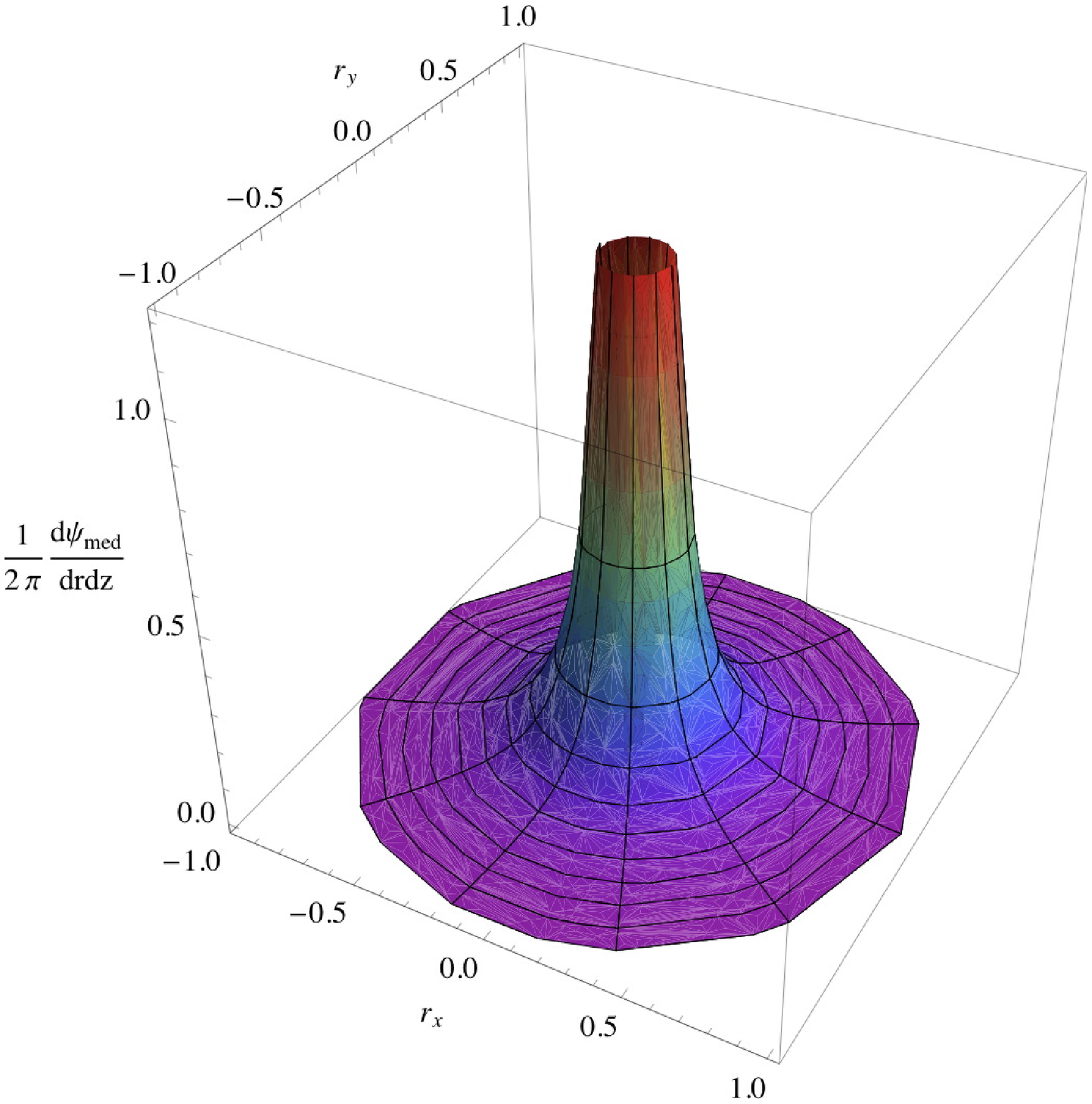,height=2.5in,clip=,angle=0}\\
\caption{(Color online) 3D plot of the double differential
medium-induced jet shapes for a quark jet of jet energy $E_T=100$~GeV
with $R=1,\, \omega^{\min}=0$~GeV in Pb+Pb collisions with
$\sqrt{s}=5500$~GeV at the LHC. Four different figures represent
the double differential jet shape for medium-induced gluon
momentum fraction $z = 0.01,\, 0.03, \, 0.1, \, 0.3 $, respectively.
}
\label{fig:LHCshapes-Medium-3D}
\end{figure*}

In this paper we have investigated extensively the differential jet
shapes $\psi(r,R) = \frac{d\Psi_{\rm int.}(r/R)}{dr}$ in vacuum and in
heavy-ion collisions at the LHC. This quantity integrates over the
energy
distribution of the partonic jet fragments. Thus, the information about
the angular distribution of soft vs hard shower partons is lost. At the
LHC it might be possible, via particle tracking or jet re-analysis, to
recover this information on an event-by-event basis and construct
$\frac{d^2\Psi(r,R)}{drdz}$, where:
\begin{equation}
   \frac{d^2\Psi_{\rm med.}(r,R)}{drdz}
= \frac{1}{\Delta E^{in}(R,0)}\frac{dI^g(\omega=zE_{jet},r)}{d\omega
dr} \;.
\label{dpsidrdz}
\end{equation}
In Eq.~(\ref{dpsidrdz}) the QGP-induced double differential shape is
normalized such that it integrates to unity with $R^{\max}=R$,
$\omega^{\min} = 0$ and $z = \omega/E_{jet}$. To illustrate the
additional
insight that can be gained through such studies we show numerical
results for a simplified case where a quark jet of $E_{jet}=100$~GeV
propagating through a QGP of length $L=6$~fm. While the medium is not
expanding and characterized by $\lambda_g = 1.5$~fm, $m_D = 0.7$~GeV
and $\alpha_s=0.3$, the $\langle \Delta E / E \rangle$ approximates
the full numerical result for central Pb+Pb collisions at the LHC.

Fig.~\ref{fig:LHCshapes-Medium-3D} shows a 3D plot of
  $ r \frac{\psi_{med}(r,R)}{(2\pi r)drdz } $ for momentum fraction
$z = k^+/E^+ \sim \omega_{gluon}/E_T = 0.01,\, 0.03,\, 0.1,\, 0.3$
and $R=1$, $\omega^{\min}=0$~GeV. We can observe that at $z=0.01$
($\omega = 1$~GeV) the double differential medium-induced jet shape
is dominated by gluon radiation at large opening angle $r/R$. At
$z=0.03$
($\omega = 3$~GeV) the peak  lies in the intermediate $r/R$ region with
significantly suppressed gluon radiation at small open angle $r/R$.
Increasing $\omega$ further narrows the medium-induced intensity
profile. It is tempting to associate the characteristic shapes for small
values of $z$ in  Fig.~\ref{fig:LHCshapes-Medium-3D} with
measurements of enhanced  away-side large-angle particle-triggered
correlations~\cite{:2008cq,Ulery:2006ha,Molnar:2008jg}. However, the
current RHIC data presents challenges in separating the jet from
the background or even distinguishing between events with 1 or 2
minijets
(recall that $N_{coll} \sim 1000$ in central Au+Au). Analysis on an
event-by-event basis can help reveal unambiguously the QGP medium
response to jets. When future experimental measurements of jet in
heavy ion collisions are perfected at the LHC full numerical simulations
of the double differential shape, including the vacuum and medium-
induced
components, will soon follow.

\end{appendix}



\begin{thebibliography}{99}


\bibitem{Wang:1991xy}
  X.~N.~Wang and M.~Gyulassy,
  Phys.\ Rev.\ Lett.\  {\bf 68}, 1480 (1992).


\bibitem{Gyulassy:2003mc}
  M.~Gyulassy, I.~Vitev, X.~N.~Wang and B.~W.~Zhang,
  arXiv:nucl-th/0302077.


\bibitem{:2008cx}
  A.~Adare {\it et al.}  [PHENIX Collaboration],
  arXiv:0801.4555 [nucl-ex].


\bibitem{Aggarwal:2007gw}
  M.~M.~Aggarwal {\it et al.}  [WA98 Collaboration],
  Phys.\ Rev.\ Lett.\  {\bf 100}, 242301 (2008)
  [arXiv:0708.2630 [nucl-ex]].



\bibitem{BDMPS-Z-ASW}
  R.~Baier, Y.~L.~Dokshitzer, A.~H.~Mueller, S.~Peigne and D.~Schiff,
  Nucl.\ Phys.\  B {\bf 484}, 265 (1997)
  [arXiv:hep-ph/9608322];
  B.~G.~Zakharov,
  JETP Lett.\  {\bf 73}, 49 (2001)
  [Pisma Zh.\ Eksp.\ Teor.\ Fiz.\  {\bf 73}, 55 (2001)]
   [arXiv:hep-ph/0012360].
  N.~Armesto, C.~A.~Salgado and U.~A.~Wiedemann,
  Phys.\ Rev.\  D {\bf 69}, 114003 (2004)
  [arXiv:hep-ph/0312106].


\bibitem{GLV}
  M.~Gyulassy, P.~Levai and I.~Vitev,
  Phys.\ Rev.\ Lett.\  {\bf 85}, 5535 (2000)
  [arXiv:nucl-th/0005032];
  I.~Vitev,
  Phys.\ Rev.\  C {\bf 75}, 064906 (2007)
  [arXiv:hep-ph/0703002].



\bibitem{HT}
  X.~N.~Wang and X.~F.~Guo,
  Nucl.\ Phys.\  A {\bf 696}, 788 (2001)
  [arXiv:hep-ph/0102230];
  B.~W.~Zhang and X.~N.~Wang,
  Nucl.\ Phys.\  A {\bf 720}, 429 (2003)
  [arXiv:hep-ph/0301195].


\bibitem{AMY}
  P.~Arnold, G.~D.~Moore and L.~G.~Yaffe,
  JHEP {\bf 0206}, 030 (2002)
  [arXiv:hep-ph/0204343].


\bibitem{Wicks:2008ta}
  S.~Wicks,
  arXiv:0804.4704 [nucl-th].


\bibitem{Adare:2008cg}
  A.~Adare {\it et al.}  [PHENIX Collaboration],
  Phys.\ Rev.\  C {\bf 77}, 064907 (2008)
  [arXiv:0801.1665 [nucl-ex]].

\bibitem{Bass:2008rv}
  S.~A.~Bass, C.~Gale, A.~Majumder, C.~Nonaka, G.~Y.~Qin, T.~Renk and J.~Ruppert,
  arXiv:0808.0908 [nucl-th].


\bibitem{Collins:2007nk}
  J.~Collins and J.~W.~Qiu,
  Phys.\ Rev.\  D {\bf 75}, 114014 (2007)
  [arXiv:0705.2141 [hep-ph]].


\bibitem{D'Enterria:2007xr}
  D.~G.~.~d'Enterria {\it et al.}  [CMS Collaboration],
  J.\ Phys.\ G {\bf 34}, 2307 (2007).
  N.~Grau [ATLAS Collaboration],
  J.\ Phys.\ G {\bf 35}, 104040 (2008)
  [arXiv:0805.4656 [nucl-ex]].


\bibitem{Salur:2008hs}
  S.~Salur  [STAR Collaboration],
  arXiv:0809.1609 [nucl-ex].


\bibitem{Seymour:1997kj}
  M.~H.~Seymour,
  Nucl.\ Phys.\  B {\bf 513}, 269 (1998)
 [arXiv:hep-ph/9707338];
     M.~H.~Seymour,
  arXiv:hep-ph/9707349.


\bibitem{Ellis:2007ib}
  S.~D.~Ellis, J.~Huston, K.~Hatakeyama, P.~Loch and M.~Tonnesmann,
  Prog.\ Part.\ Nucl.\ Phys.\  {\bf 60}, 484 (2008)
  [arXiv:0712.2447 [hep-ph]].


\bibitem{Salgado:2003rv}
  C.~A.~Salgado and U.~A.~Wiedemann,
  Phys.\ Rev.\ Lett.\  {\bf 93}, 042301 (2004)
  [arXiv:hep-ph/0310079].


\bibitem{Dokshitzer:1995zt}
  Y.~L.~Dokshitzer and B.~R.~Webber,
  Phys.\ Lett.\  B {\bf 352}, 451 (1995)
  [arXiv:hep-ph/9504219].

\bibitem{Fischer:2002hna}
  C.~S.~Fischer and R.~Alkofer,
  Phys.\ Lett.\  B {\bf 536}, 177 (2002)
  [arXiv:hep-ph/0202202].



\bibitem{Acosta:2005ix}
  D.~E.~Acosta {\it et al.}  [CDF Collaboration],
  Phys.\ Rev.\  D {\bf 71}, 112002 (2005)
  [arXiv:hep-ex/0505013].

\bibitem{Vitev:2008jh}
  I.~Vitev,
  J.\ Phys.\ G {\bf 35}, 104011 (2008)
  [arXiv:0806.0003 [hep-ph]].


\bibitem{Vitev:2005yg}
  I.~Vitev,
  Phys.\ Lett.\  B {\bf 630}, 78 (2005)
  [arXiv:hep-ph/0501255].


\bibitem{Djordjevic:2003zk}
  M.~Djordjevic and M.~Gyulassy,
  Nucl.\ Phys.\  A {\bf 733}, 265 (2004)
  [arXiv:nucl-th/0310076];
  B.~W.~Zhang, E.~Wang and X.~N.~Wang,
  Phys.\ Rev.\ Lett.\  {\bf 93}, 072301 (2004)
  [arXiv:nucl-th/0309040].


\bibitem{Gyulassy:2001kr}
   M.~Gyulassy, I.~Vitev, X.~N.~Wang and P.~Huovinen,
   Phys.\ Lett.\  B {\bf 526}, 301 (2002)
   [arXiv:nucl-th/0109063].

\bibitem{Sjostrand:2006za}
  T.~Sjostrand, S.~Mrenna and P.~Skands,
  JHEP {\bf 0605}, 026 (2006)
  [arXiv:hep-ph/0603175].

\bibitem{Markert:2008jc}
   C.~Markert, R.~Bellwied and I.~Vitev,
   arXiv:0807.1509 [nucl-th].

\bibitem{Vitev:2005he}
  I.~Vitev,
  Phys.\ Lett.\  B {\bf 639}, 38 (2006)
  [arXiv:hep-ph/0603010].

\bibitem{Majumder:2007iu}
  A.~Majumder,
  J.\ Phys.\ G {\bf 34}, S377 (2007)
  [arXiv:nucl-th/0702066].

\bibitem{Wicks:2007zz}
  S.~Wicks and M.~Gyulassy,
  J.\ Phys.\ G {\bf 34}, S989 (2007)
  [arXiv:nucl-th/0701088].

\bibitem{Adil:2006ra}
  A.~Adil and I.~Vitev,
  Phys.\ Lett.\  B {\bf 649}, 139 (2007)
  [arXiv:hep-ph/0611109].


\bibitem{Lokhtin:2007ga}
  I.~P.~Lokhtin, S.~V.~Petrushanko, A.~M.~Snigirev and C.~Y.~Teplov,
  PoS {\bf LHC07}, 003 (2007)
  [arXiv:0706.0665 [hep-ph]].
  I.~P.~Lokhtin, L.~V.~Malinina, S.~V.~Petrushanko, A.~M.~Snigirev, I.~Arsene and K.~Tywoniuk,
  arXiv:0809.2708 [hep-ph].


\bibitem{Vitev:2008vk}
  I.~Vitev and B.~W.~Zhang,
  arXiv:0804.3805 [hep-ph].



\bibitem{Mironov:2007bu}
  C.~Mironov, P.~Constantin and G.~J.~Kunde,
  Eur.\ Phys.\ J.\  C {\bf 49}, 19 (2007).


\bibitem{Grau:2008ed}
  N.~Grau, B.~A.~Cole, W.~G.~Holzmann, M.~Spousta and P.~Steinberg,
  arXiv:0810.1219 [nucl-ex].




\bibitem{Vitev:2006bi}
  I.~Vitev, J.~T.~Goldman, M.~B.~Johnson and J.~W.~Qiu,
  Phys.\ Rev.\  D {\bf 74}, 054010 (2006)
  [arXiv:hep-ph/0605200].

\bibitem{Pumplin:2002vw}
  J.~Pumplin, D.~R.~Stump, J.~Huston, H.~L.~Lai, P.~Nadolsky and W.~K.~Tung,
  JHEP {\bf 0207}, 012 (2002)
  [arXiv:hep-ph/0201195].

\bibitem{Abulencia:2005jw}
  A.~Abulencia {\it et al.}  [CDF II Collaboration],
  Phys.\ Rev.\ Lett.\  {\bf 96}, 122001 (2006)
  [arXiv:hep-ex/0512062].




\bibitem{Klasen:1997tj}
  M.~Klasen and G.~Kramer,
  Phys.\ Rev.\  D {\bf 56}, 2702 (1997)
  [arXiv:hep-ph/9701247].



\bibitem{:2008cq}
   A.~Adare {\it et al.}  [PHENIX Collaboration],
   Phys.\ Rev.\  C {\bf 78}, 014901 (2008)
   [arXiv:0801.4545 [nucl-ex]];
   J.~Jia,
   arXiv:0805.0160 [nucl-ex].

\bibitem{Ulery:2006ha}
   J.~G.~Ulery  [STAR Collaboration],
   Nucl.\ Phys.\  A {\bf 783}, 511 (2007)
   [arXiv:nucl-ex/0609047];

   C.~A.~Pruneau  [STAR Collaboration],
   J.\ Phys.\ G {\bf 34}, S667 (2007).


\bibitem{Molnar:2008jg}
   L.~Molnar,
   PoS {\bf LHC07}, 027 (2007)
   [arXiv:0801.2715 [nucl-ex]].


\end{thebibliography}
\end{document}